\documentclass[11pt]{iopart}
\usepackage{iopams} 
\pdfoutput=1

\usepackage{graphicx}
\expandafter\let\csname equation*\endcsname\relax

\expandafter\let\csname endequation*\endcsname\relax

\usepackage{amsmath,amsfonts,amsthm,bm}

\usepackage{stackengine}
\newcommand\mathstack[2]{\ensurestackMath{%
  \stackengine{\Lstackgap}{{}#1{}}{#2}{O}{c}{F}{F}{L}}}
\usepackage{cite} 

\usepackage{rotate}
\usepackage{color}

\DeclareFontFamily{OT1}{pzc}{}
\DeclareFontShape{OT1}{pzc}{m}{it}%
            {<-> s * [1.10] pzcmi7t}{}
\DeclareMathAlphabet{\mathscr}{OT1}{pzc}%
                                {m}{it}

\definecolor{RedWine}{rgb}{0.743,0,0}
\definecolor{RoyalBlue}{rgb}{0.25,.41,.88}
\definecolor{ForestGreen}{rgb}{.13,.54,.13}
\usepackage[backref,breaklinks,colorlinks,citecolor=ForestGreen]{hyperref}

\newcommand{\be}{\begin{equation}}
\newcommand{\ee}{\end{equation}}
\newcommand{\bea}{\begin{eqnarray}}
\newcommand{\eea}{\end{eqnarray}}
\def\ba#1\ea{\begin{align}#1\end{align}}

\begin{document}

\title{Self-force effects in post-Minkowskian scattering}

\author{Samuel E. Gralla and Kunal Lobo}

\address{Department of Physics, University of Arizona, Tucson, Arizona 85721, USA}

\ead{klobo@email.arizona.edu}

\begin{abstract}
We revisit the old problem of the self-force on a particle moving in a weak-field spacetime in the context of renewed interest in two-body gravitational scattering.  We analytically calculate the scalar, electromagnetic, and gravitational self-force on a particle moving on a straight-line trajectory at a large distance from a Newtonian star, and use these results to find the associated correction to its motion.  In the gravitational case we must also include the matter-mediated force, which acts at the same perturbative order as the gravitational self-force.  We further augment the gravitational results with geodesic calculations at second order in the central body mass to determine the full, explicit solution to the two-body gravitational scattering problem at second post-Minkowskian order (2PM).  We calculate the momentum transfer (which reproduces Westpfahl's old result), the change in mechanical angular momentum (which matches the radiative flux recently computed by Damour), and the change in mechanical mass moment (the time-space components of the angular momentum tensor), which has not previously appeared.  Besides the new 2PM results of explicit trajectories and all conserved quantities, this work clarifies the role of gravitational self-force in PM scattering theory and provides a foundation for higher-order calculations.
\end{abstract}

\maketitle

\section{Introduction}

The study of self-force effects in curved spacetime began in 1960 with DeWitt and Brehme's foundational paper deriving the ``tail integral'' formula for electromagnetic self-force \cite{DeWitt:1960fc}.  A few years later, DeWitt and DeWitt \cite{DeWitt:1964de} evaluated this integral in the leading post-Newtonian (PN) approximation, i.e., for a charged particle moving slowly in the weak gravitational field of a point mass.  This calculation was extended to the leading post-\textit{Minkowskian} (PM) approximation (small-angle scattering with arbitrary initial velocity) by Westpfahl and Goller in 1980 \cite{westpfahl1980relativistic} and the results were incorporated into Westpfahl's 1985 treatise on relativistic scattering \cite{Westpfahl:1985tsl}.  These early PN and PM calculations illustrated the interesting physical effects of the self-force, but were (to quote the original) of ``conceptual interest only, since the forces involved are far too small to be detected experimentally'' \cite{DeWitt:1960fc}.

In the 1990's new interest in the self-force emerged with the realization that \textit{gravitational} self-force effects may in fact be of \textit{practical} interest for gravitational-wave astronomy.  The equations of gravitational self-force were formulated \cite{Mino:1996nk,Quinn:1996am} and it was realized that a scalar  toy model \cite{Quinn:2000wa} would provide a simpler starting point for computations.  In light of these developments, Pfenning and Poisson \cite{Pfenning:2000zf} revisited the DeWitt-DeWitt PN calculation, clarifying the electromagnetic case and deriving analogous scalar and gravitational results.  They illustrated the necessity of including additional ``matter-mediated'' forces in a consistent treatment of binary systems and connected the self-force method with the standard PN approach.  Pfenning and Poisson's calculations clarified the role of the self-force formalism in binary dynamics and---in our view---were invaluable in establishing context for its more ambitious goal of providing accurate gravitational waveforms for relativistic binaries  \cite{Barack:2009ux}.

Motivated by the realization that gravitational \textit{scattering} provides important information about binary dynamics and has deep connections to quantum processes and methods \cite{Damour:2016gwp,Damour:2017zjx,Bjerrum-Bohr:2018xdl,Cheung:2018wkq,Kosower:2018adc,Bern:2019nnu,Bern:2019crd,Damour:2019lcq,Kalin:2019rwq,Kalin:2019inp,Kalin:2020fhe,Kalin:2020mvi,Mogull:2020sak,DiVecchia:2020ymx,Cheung:2020gbf,Bjerrum-Bohr:2021vuf,Jakobsen:2021smu,Bern:2021dqo,Herrmann:2021tct,Herrmann:2021lqe,DiVecchia:2021bdo,Dlapa:2021npj,Bini:2021gat,Brandhuber:2021eyq}, in this paper we will analogously revisit the Westpfahl-Goller PM self-force calculations.  We consider a particle moving on a nearly-straight trajectory at a great distance from a Newtonian star and calculate the leading effects of its (scalar, electromagnetic, or gravitational) self-field.  We rederive the Westfpahl-Goller electromagnetic self-force and provide analogous scalar and gravitational results.  For the gravitational case, we review the necessity of including the matter-mediated force and calculate it in closed form.  We explicitly integrate the perturbed equations of motion in all three cases, providing parameterized trajectories in terms of elementary functions.

These results give the motion of the particle (mass $m$) in the initial rest frame of the star (mass $M$), meaning the frame where the star has asymptotically zero velocity in the infinite past.  For gravitational scattering, we also determine the motion of the star by invoking the mass exchange symmetry of the problem.  This requires augmenting our $O(Mm)$ self-force calculation with test-mass results at order $O(M^2)$ and changing to the center of energy-momentum (CEM) frame, where the symmetry is manifest as $M \leftrightarrow m$ together with rotation by $\pi$.  This procedure yields the full parameterized trajectories of both bodies in 2PM scattering.  

While the 2PM scattering problem has been studied by many different groups over the years, it appears that the full trajectories are a new result.  From these trajectories we may compute any mechanical property of 2PM scattering.  We rederive the momentum transfer (scattering angle) first computed by Westpfahl \cite{Westpfahl:1985tsl}.  We compute the change in mechanical angular momentum, which matches the radiative angular momentum flux calculated by Damour \cite{Damour:2020tta}.  We also compute the change in mechanical mass moment (the time-space component of the relativistic angular momentum tensor), a new result.  We will be exploring further aspects of this change in mass moment in future publications \cite{gravscoot,EMscoot}.

This paper is organized as follows.  We begin with a prelude on the scattering angle that places our work in the context of the extensive recent interest in this quantity (Sec.~\ref{sec:prelude}).  The remaining sections then derive the results, as follows.  In Sec.~\ref{sec:green} we review the construction of the scalar, electromagnetic, and gravitational retarded Green functions in a weakly curved spacetime at $O(M)$.  In Sec.~\ref{sec:self-forces} we evaluate the tail integral to derive the self-forces at this order in the three cases.  In Sec.~\ref{sec:matter-mediated} we derive the matter-mediated force that must be included in the gravitational case.  In Sec.~\ref{sec:motion} we integrate the equations of motion and discuss physical quantities defined in the star frame.  In Sec.~\ref{sec:PM} we add higher-order geodesic calculations and change to the CEM frame, providing the full 2PM trajectories and calculating associated physical quantities.  Our metric signature is $-+++$ and we use Gaussian units with $G=c=1$.  Covariant derivatives are denoted with a $\nabla$ or a semicolon, while partial derivatives are denoted with a $\partial$ or a comma.  Symmetrization is denoted with parentheses, e.g., $T_{(\alpha \beta)} = (1/2) (T_{\alpha \beta}+T_{\beta \alpha})$.

\section{Prelude: scattering angle}\label{sec:prelude}

The PM scattering angle has been the subject of intensive interest since Damour's 2016 analysis of its close relationship to the conservative dynamics of bound systems \cite{Damour:2016gwp}.  In order to place our work in the context of this active area of research, we now describe our approach through the lens of the scattering angle.  Our calculation is organized in a joint perturbation series in the mass $M$ of the star and the ``charge'' $Q$ of the body, taken to be either the scalar charge $q$, the electric charge $e$, or the mass $m$.  The relevant dimensionless parameters are 
\begin{align}
    \frac{M}{bv^2} \ll 1, \qquad \frac{Q^2}{mb} \ll 1,
\end{align}
where $b$ is the impact parameter and $v$ is the initial relative velocity.  We take the star to be at rest at zeroth order, and  compute the leading self-force effects on the particle by evaluating the tail integral along straight line motion in the linearized Schwarzschild metric.  In the gravitational case we must additionally solve for the $O(m/b)$ motion of the star and take into account the corrected gravitational forces on the particle---the so-called matter-mediated forces introduced by \cite{Pfenning:2000zf}.  Integrating the equations gives the motion of the particle in the ``star frame'', meaning the frame where the star has asymptotically zero velocity in the distant past.

For the particle's scattering angle $\delta$ in the star frame, we find
\begin{align}
    \delta_{\rm scalar} & = \frac{M}{bv^2}\left(2(1+v^2)- \frac{ q^2}{mb}\frac{\pi}{4}(v^2+4\xi(1-v^2)) + \dots \right) + O\left( \frac{M^2}{b^2} \right) \label{deltascalar_intro} \\
    \delta_{\rm em} & = \frac{M}{bv^2}\left(2(1+v^2)-\frac{e^2}{mb}\frac{\pi}{4}(v^2+2) + \dots \right) + O\left( \frac{M^2}{b^2} \right) \label{deltaem_intro} \\
    \delta_{\rm grav} & = \frac{M}{bv^2}\left(2(1+v^2) + \frac{m}{b} \frac{3\pi}{4}(v^2+4) + \dots \right) + O\left( \frac{M^2}{b^2} \right),\label{deltagravity_intro}
\end{align}
where $\xi$ is the scalar coupling to curvature (see Eq.~\eqref{Gscalar} below).  To obtain the 2PM CEM-frame scattering angle $\chi$, we first  augment the $O(mM)$ result of \eqref{deltagravity_intro} with an $O(M^2)$ test mass calculation.  This geodesic contribution to the deflection $\delta_{\rm grav}$ turns out to have precisely the same numerical coefficient as the $O(Mm)$ term we computed:
\begin{align}
    \delta_{\rm grav} & = \frac{M}{bv^2}\left(2(1+v^2) + \frac{m}{b} \frac{3\pi}{4}(v^2+4) + O\left( \frac{m^2}{b^2} \right) \right) + \frac{M^2}{b^2 v^2} \frac{3\pi}{4}(v^2+4) + O\left( \frac{M^3}{b^3} \right).\label{deltagravity2_intro}
\end{align}
Eq.~\eqref{deltagravity2_intro} is the deflection angle of the particle as measured in the star frame.  Denoting the CEM-frame deflection angle by $\chi$, from a simple boost we find  $\chi=(\tilde{E}/M) \delta$ at this order of approximation, where $\tilde{E}=\sqrt{M^2+m^2+2 M m \gamma}$ is the initial total energy in the initial CEM frame (with $\gamma=(1-v^2)^{-1/2}$ the initial Lorentz factor).  We therefore derive
\begin{align}\label{chi_intro}
    \chi & = \frac{\tilde{E}}{bv^2}\left(2(1+v^2) + \left(\frac{m}{b}+ \frac{M}{b} \right) \frac{3\pi}{4}(v^2+4) \right),
\end{align}
which is consistent to 2PM in the sense that it contains all terms that scale as $\lambda^2$ under $M \to \lambda M$ and $m \to \lambda m$.  The parameter $v$ is now interpreted as the relative velocity.

At this stage $\chi$ is the CEM-frame deflection angle \textit{of the particle} (mass $m$).  However, as the CEM frame is invariant under exchange of the two bodies, the CEM-frame deflection angle \textit{of the star} (mass $M$) is determined by simply sending $m \leftrightarrow M$ in Eq.~\eqref{chi_intro}.  But this formula is symmetric under the exchange, and we conclude that the star deflects by the same angle as the particle.  We may therefore speak of \textit{the} CEM-frame deflection angle at 2PM, given by Eq.~\eqref{chi_intro}.  This reproduces Westpfahl's result \cite{Westpfahl:1985tsl}.

It is worth emphasizing the logical role played by the exchange symmetry.  The symmetry implies that $\chi_M = \chi_m(M\leftrightarrow m)$, where $\chi_M$ is the CEM-frame deflection angle of the mass $M$ (the ``star'') and $\chi_m$ is the CEM-frame deflection angle of the mass $m$ (the ``particle'').  In our approach, we \textit{derive} that $\chi_M=\chi_m$ via direct calculation, resulting from the fact that the same coefficient $3\pi(v^2+4)/4$ appears in a self-force calculation and in a geodesic calculation.  If one is willing to instead \textit{assume} that $\chi_M=\chi_m$, then the agreement of these coefficients is guaranteed, and one can predict the self-force result from a geodesic calculation. Damour \cite{Damour:2019lcq} has made this observation in the context of the  ``conservative dynamics'' at all orders, where the two bodies by definition deflect by the same amount.  While this is an enormously useful trick to obtain a portion of the dynamics from simple calculations, we emphasize that, without an independent argument that $\chi_M=\chi_m$, geodesic calculations alone cannot derive the scattering angle at 2PM or higher.

In this section we have given a preview of some results in a manner that illustrates key features of the approach.  Our ensuing derivation of the complete 2PM trajectories follows a similar pattern: First, we first obtain the $O(M m/b^2)$ trajectory of the particle in the star frame.  Next we add in the $O(M^2/b^2)$ geodesic corrections.  Finally, we calculate the particle trajectory in the the CEM frame and invoke the mass exchange symmetry to determine the trajectory of the star.  We now describe these results.

\section{Green functions in a weakly curved spacetime}\label{sec:green}

In this section we review the construction of scalar, electromagnetic, and gravitational Green functions in a weakly curved spacetime.  In any spacetime $g_{\mu \nu}$, these Green functions are defined by \cite{Pfenning:2000zf}
\begin{align}
    \Box G - \xi R G & = -4\pi \delta_4(x,x'), \label{Gscalar} \\
    \Box G^\alpha{}_{\alpha'} - R^\alpha{}_\beta  G^\beta{}_{\alpha'} & = - 4\pi \delta^\alpha{}_{\alpha'} \delta_4(x,x') \\
    \Box \bar{G}^{\alpha \beta}{}_{\alpha'\beta'} +  \mathcal{R}^\alpha{}_\gamma{}^\beta{}_\delta \bar{G}^{\gamma \delta}{}_{\alpha'\beta'} & = - 4\pi \delta^{(\alpha}{}_{\alpha'} \delta^{\beta)}{}_{\beta'} \delta_4(x,x')\label{Ggravity}
\end{align}
where $R$ is the Ricci scalar, $R_{\mu \nu}$ is the Ricci tensor, $R_{\mu \nu \alpha \beta}$ is the Riemann tensor, and
\begin{align}
    \mathcal{R}_{\mu \alpha \nu \beta} = 2 R_{\mu \alpha \nu \beta} + 2 R_{\mu ( \alpha}g_{\beta) \nu} - R_{\mu \nu} g_{\alpha \beta} - 2 R_{\alpha \beta} g_{\mu \nu} - R g_{\mu(\alpha}g_{\beta)\nu} + R g_{\mu \nu} g_{\alpha \beta}.
\end{align}
We have written  $\Box=g^{\alpha \beta}\nabla_\alpha \nabla_\beta$ for the wave operator and $\delta_4(x,x') = \delta^{(4)}(x-x')/\sqrt{-g}$ for the invariant Dirac delta function.  The scalar Green function $G$ is a Green function for a massless scalar field with curvature coupling $\xi$.  The electromagnetic Green function $G^\alpha{}_{\alpha'}$ is a Green function for the gauge field $A^\mu$ in Lorenz gauge, $\nabla_\mu A^\mu=0$.  The gravitational Green function $\bar{G}^{\alpha \beta}{}_{\alpha' \beta'}$ is a Green function for the trace-reversed metric perturbation $\bar{h}_{\alpha \beta}=h_{\alpha \beta}-(1/2) h^\mu{}_\mu g_{\alpha \beta}$ in Lorenz gauge, $\nabla^\alpha \bar{h}_{\alpha \beta}=0$.  A Green function for the metric perturbation $h_{\mu \nu}$ is given by
\begin{align}
    G^{\alpha \beta}{}_{\alpha' \beta'} & = \bar{G}^{\alpha \beta}{}_{\alpha' \beta'} - \frac{1}{2} g^{\alpha \beta}g_{\gamma \delta} \bar{G}^{\gamma \delta}{}_{\alpha' \beta'}.
\end{align}
Our normalization for the gravitational Green function follows Ref.~\cite{Pfenning:2000zf}, differing from the conventional one by a factor of four.  (In particular, the integral $\int G T$ gives one-quarter the trace-reversed metric perturbation.) In all cases we consider the retarded Green function, i.e., the solution that vanishes when $x'$ is not in the causal past of $x$.  

We will construct these Green functions in a weakly curved spacetime following the approach of Pfenning and Poisson \cite{Pfenning:2000zf}.  The spacetime is described by a static Newtonian potential $\Phi(x,y,z)$, with metric 
\begin{align}\label{spacetime}
ds^2 = -(1+2\Phi)dt^2 + (1-2\Phi)(dx^2 + dy^2 + dz^2) + O(\Phi^2).
\end{align}
The Newtonian mass density is given by Poisson's equation,
\begin{align}
    \nabla^2 \Phi = 4 \pi \rho.
\end{align}
Pfenning and Poisson \cite{Pfenning:2000zf} define three key biscalars from which all the Green functions follow by differentiation.  These are the retarded Green function in flat spacetime
\be\label{Gflat}
G_{\rm flat}(x,x')= \frac{\delta(t-t'-|\mathbf{x}-\mathbf{x}'|)}{|\mathbf{x}-\mathbf{x}'|},
\ee
together with additional biscalars $A$ and $B$ defined by 
\begin{align}
A(x,x') & = \frac{1}{2\pi}\int G_{\rm flat}(x,x'')\Phi(x'')G_{\rm flat}(x'',x')d^4 x'' \label{A} \\
B(x, x') & = \int G_{\rm flat}(x,x'')\rho(x'')G_{\rm flat}(x'',x')d^4 x''. \label{B}
\end{align}

The scalar Green function is given by Eqs.~(3.11) and (3.14) of Ref.~\cite{Pfenning:2000zf},
\begin{align} \label{Green}
G = G_{\rm flat} - 2A_{,tt'}-2\xi B + O(\Phi^2),
\end{align}
while electromagnetic and gravitational Green functions are similarly expressed in terms of $A$ and $B$ in Eqs.~(3.17), (3.21), (3.30), and (3.32).\footnote{For Eq. (3.21), the reader should refer to the arXiv version of \cite{Pfenning:2000zf}, since the journal version contains a typographical error.}

These results hold for any weak-field spacetime.  We now specialize to a star (mass $M$) that is compactly supported within some radius $\mathcal{R}$ and consider the Green functions only at distant spatial points,
\begin{align}\label{bigr}
    |\mathbf{x}|,|\mathbf{x}'| \gg \mathcal{R}.
\end{align}
The Newtonian potential at these distances can be approximated by $\Phi=-M/r$, but the integrals \eqref{A} and \eqref{B} defining the biscalars (and ultimately providing the Green functions) involve the entire spacetime, including regions where this approximation is invalid.  However, it turns out nevertheless to be consistent to make this replacement,
\begin{align}\label{sub}
    \Phi \to - \frac{M}{r}, \qquad \rho \to M \delta (\mathbf{x}),
\end{align}
after which the integrals yield \cite{DeWitt:1960fc,Pfenning:2000zf}
\begin{align}
    A & = -\frac{M}{R}\Theta(T-R)
    \begin{cases} \displaystyle
    \log  \frac{r + r' + R}{r+r'-R} &  T < r+r' \vspace{2mm} \\ 
    \displaystyle \log \frac{T+R}{T-R} & T > r+r',
    \end{cases}\label{A2} \\
    B & = \frac{M}{r r'} \delta(T-r-r'),\label{B2}
\end{align}
where $r=|\bm{x}|$, $r'=|\bm{x}'|$, $R=|\bm{x}-\bm{x'}|$, and $T=t-t'$.  Note that with the substitution $\Phi=-M/r$, the metric \eqref{spacetime} agrees with the Schwarzschild metric in isotropic coordinates, expanded to first order in $M$.

We now justify the substitution \eqref{sub} under the approximation \eqref{bigr}.  For the $B$ integral, we will rely on the careful arguments of Sec.~IVD of Ref.~\cite{Pfenning:2000zf}, which derive Eq.~\eqref{B2} directly.   For the $A$ integral, we note that the portion of the integration region where $\Phi$ differs significantly from $-M/r$ is negligible in the approximation \eqref{bigr}.  This may be visualized noting that the the integration region in \eqref{A} (and \eqref{B}) may be identified with the ellipsoid $(\bm{x}''-\bm{x})^2+(\bm{x}''-\bm{x}')^2=(t-t')^2$ in Euclidean space parameterized by $\bm{x}''$, which has foci at the two spatial points $\bm{x}'$ and $\bm{x}''$.  As these two foci are by assumption located at large distances from the region of the star (Eq.~\eqref{bigr}), that region occupies only a parametrically small portion of the ellipsoid, which can be neglected in the integral at leading order.  Formally, one may approximate the integral using matched asymptotic expansions with small parameter $b/\mathcal{R}$, where $b$ is the greater of $r$ and $r'$, defining a near-zone near-zone $r'' \ll b$ and a far-zone $r'' \gg \mathcal{R}$.  One finds that the near-zone contribution vanishes at leading order in $b/\mathcal{R}$.

We now illustrate the properties of the resulting Green functions, using the scalar case as an example.  Plugging Eqs.~\eqref{Gflat}, \eqref{A2} and \eqref{B2} in to Eq.~\eqref{Green} and dropping the $O(\Phi^2)$ error, one finds
\begin{align}
    G = G^{\rm direct} + G^{\rm tail}
\end{align}
with
\begin{align}
    G^{\rm direct} & = \frac{\delta(T-R)}{R} - \frac{2M}{R}\log \frac{r+r'+ R}{r+r'-R} \delta'(T-R) \label{Gdirect} \\
    G^{\rm tail} & = \left( \frac{4M}{T^2-R^2} - \xi \frac{2M}{r r'}\right) \delta(T-r-r') - \frac{8MT}{(T^2-R^2)^2}\Theta(T-r-r'). \label{Gtail}
\end{align}
We have grouped the terms into the ``direct'' and ``tail'' pieces in the Hadamard decomposition (e.g., \cite{Poisson:2011nh}).

The direct piece of the Green function by definition has support only on the past light cone (i.e., when a future-directed null geodesic runs from $x'$ to $x$).  Here we see this property reflected in a series expansion in $M$; the first term involves the flat spacetime past light cone $T=R$, while the term proportional to $M$ gives the first correction.  Consistent to this perturbative order, the direct term can equivalently be written
\begin{align}
    G^{\rm direct} =  \frac{1}{R}\delta(\Sigma), \qquad \Sigma = T - R - 2 M\log \frac{r+r'+ R}{r+r'-R},
\end{align}
where $\Sigma=0$ describes the past light cone of $x$.   On general grounds, it is also possible to express this term as $G^{\rm direct}= \sqrt{\Delta} \Theta(T) \delta (\sigma)$, where $\Delta$ is the Van Vleck determinant and $\sigma$ is Synge's world function (one-half the squared geodesic distance) \cite{Poisson:2011nh}.  This form of the Green function was explored in Ref.~\cite{Chu:2011ip}.  

\begin{figure}
    \centering
    \includegraphics[scale=.3]{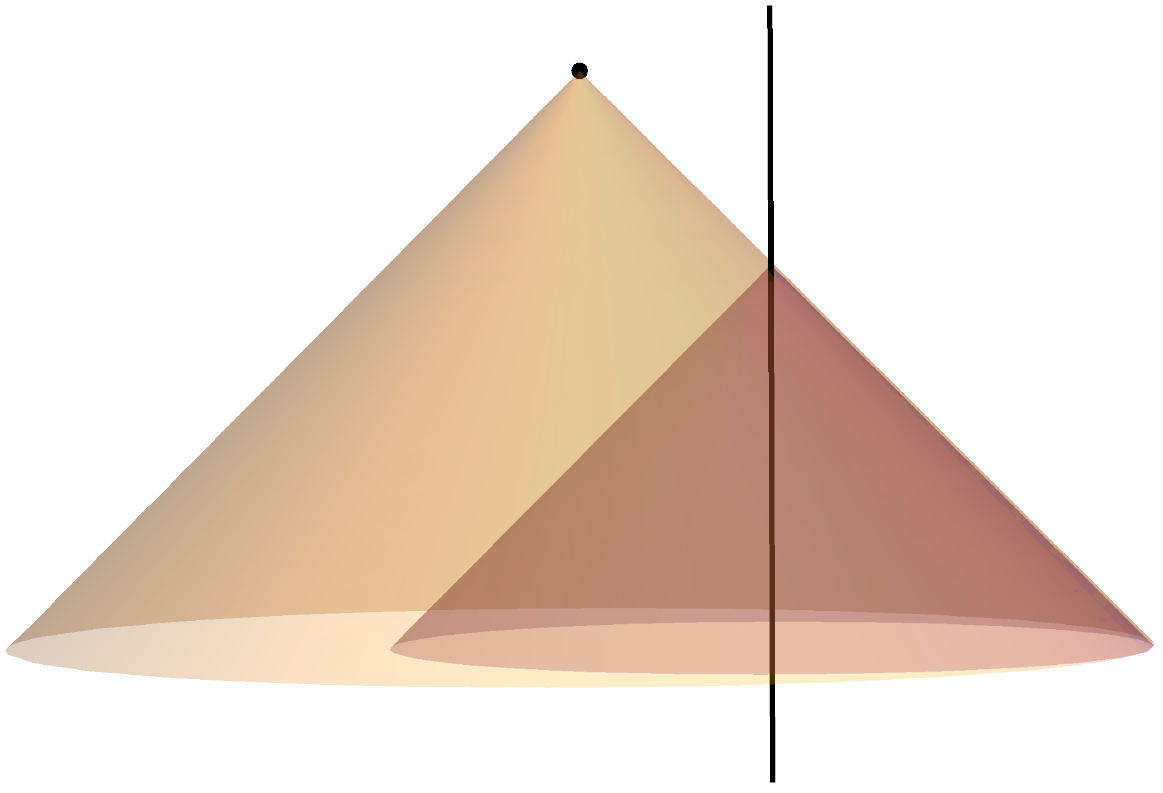}
    \caption{Support of the retarded Green function (scalar, electromagnetic, or gravitational) in the spacetime of a Newtonian star, when both points are at large spatial distances from the star.  The star can be idealized as occupying a worldline at the spatial origin of coordinates, shown as a vertical line.  The direct portion of the Green function has support on the past light cone, which can be represented as delta functions and derivatives on the flat-spacetime past light cone $T=R$ (gold).  The tail portion of the Green function has support only on and within a secondary cone $T=r+r'$ emanating from the intersection of the past light cone with the star's worldline (beige).}
    \label{fig:GF}
\end{figure}
The tail portion of the Green function has support within the past light cone.  Generically this support extends throughout the past light cone, but in the weakly curved spacetime when both points of Green function are at large distances from the star, the support is in fact only within a restricted region of the past light cone (Fig.~\ref{fig:GF}).  In particular, the tail portion vanishes when the spacetime points are sufficiently close together, becoming non-zero only when a signal has had time to ``bounce'' off the star ($T\geq r+r'$).  In this approximation, therefore, the tail may be viewed as being caused by scattering of the direct part off of the singularity at $r=0$.  These properties extend to the electromagnetic and gravitational Green's functions, and in fact were first discussed in the electromagnetic case in the original work of DeWitt and DeWitt \cite{DeWitt:1960fc}.  We are unaware of a deeper explanation for this surprising behavior of the retarded Green function.

\section{Self-forces}\label{sec:self-forces}

Consider a point particle moving on a timelike geodesic $z^\mu(\tau)$ of a spacetime $g_{\mu \nu}$, parameterized by proper time $\tau$ with four-velocity $u^\mu$.  The scalar, electromagnetic, and gravitational self-forces are given by (in the conventions of Ref.~\cite{Pfenning:2000zf})
\begin{align}
    f_{(q)}^{\alpha}(\tau) & = q^2\int_{-\infty}^{\tau^-}\left( \nabla^\alpha G+ u^\alpha u^\beta \nabla_\beta G \right) d\tau ',\label{fq} \\
    f^\alpha_{(e)}(\tau) & = e^2\int_{-\infty}^{\tau^-}\left(\nabla^\alpha G_{\beta \alpha'} - \nabla_\beta G^{\alpha}{}_{\alpha'} \right)u^\beta u^{\alpha'}d\tau' \label{fe}\\
    f^\alpha_{(m)}(\tau) & = 2m^2 \int_{-\infty}^{\tau^-}\left(\nabla^\alpha G_{\beta\gamma\mu'\nu'} - 2\nabla_\gamma G^{\alpha}{}_{\beta\mu'\nu'} - u^\alpha u^\delta \nabla_\delta G_{\beta\gamma\mu'\nu'} \right) u^\beta u^\gamma u^{\mu'} u^{\nu'} d\tau'. \label{fm}
\end{align}
Here $q$ is the scalar charge, $e$ is the electric charge, and $m$ is the mass; we denote the corresponding forces with a subscript featuring the associated symbol in parentheses.  In these expressions, both spacetime points $x$ and $x'$ are evaluated on the worldline, i.e., $x=z(\tau)$ and $x'=z(\tau')$.  For the bitensors $\nabla G$ and the tensor $u$, the prime (or lack thereof) on the index indicates which spacetime point is being considered.  The upper limit $\tau^-$ indicates that the integral is to be stopped at $\tau'=\tau-\epsilon$ for some $\epsilon>0$, with the limit $\epsilon \to 0$ taken after integration.  This excludes the singular behavior of the Green function at coincidence, having the effect of picking out the tail contribution only.  

We will consider the weakly curved spacetime \eqref{spacetime} of a compact star and choose a geodesic that remains always distant from its typical radius $\mathcal{R}$.  In computing the self-force, both points in the Green function are thus always in the large-distance regime \eqref{bigr}, so we may use the definite expressions \eqref{A2} and \eqref{B2} for the biscalars underlying the Green functions.  Letting $b$ denote the impact parameter, we may summarize the key assumptions as
\begin{align}\label{approx}
    \Phi(x,y,z) \ll 1, \qquad b \gg \mathcal{R}.
\end{align}
That is, the spacetime is weakly curved everywhere, and the particle remains far from the star.

Under these approximations at first order beyond flat spacetime, the (scalar, electromagnetic, or gravitational) Green function is constructed from $A$ and $B$ given in Eqs.~\eqref{A2} and \eqref{B2} and hence is sensitive only to the total mass $M$ of the star.  Thus, all results may be expressed in terms of a single small parameter,
\begin{align}
    \frac{M}{b} \ll 1.
\end{align}
We will organize our results in terms of this parameter; however, it should be borne in mind that the more restrictive conditions \eqref{approx} must hold for the results to be valid.

To zeroth order in $M/b$, spacetime is flat and the particle geodesic is a straight line.  We will choose the motion to be in the $z$ direction, with separation from the star in the $x$ direction,
\begin{align}\label{z}
    z^\mu(\tau) 
    = (t, b, 0, v t) + O(M/b),
\end{align}
where $v>0$ is the (constant) velocity.  The four-velocity is thus
\begin{align}\label{u}
    u^\alpha = \gamma(1,0,0,v) + O(M/b),
\end{align}
where $\gamma=1/\sqrt{1-v^2}$.  

\subsection{Results}

The tail portion of the Green function is non-zero first at $O(M/b)$, meaning that the $O(M/b)$ self-force depends only on the $O(1)$ straight-line motion of the charge.  That is, for the purposes of computing the leading, $O(M/b)$ self-force, we may consistently take the particle to move on a straight line.  Our task is therefore to compute the integrals \eqref{fq}, \eqref{fe}, and \eqref{fm} for the worldline \eqref{z} and Green function constructed from \eqref{A2} and \eqref{B2}  according to Eq.~\eqref{Green} and the description below.

The $y$ component of the self-force is zero by symmetry (the motion is confined to the $xz$ plane), while the $t$ component is related to the $z$ component by $f^t=v f^z$ according to the orthogonality condition $f^\mu u_\mu=0$.  We find that the $x$ and $z$ components can be expressed in terms of nine master integrals $\{\mathcal{A}_i,\mathcal{B}_i, \mathcal{C}_i\}$ as
\begin{align}
f_{(q)}^z & = -2q^2\gamma^{-1}\left(\gamma^2v\mathcal{A}_1+ \mathcal{A}_2  +\xi\gamma^2v\mathcal{B}_1 +\xi \mathcal{B}_2 \right) \label{scalarforcez} \\
f_{(q)}^x & = -2q^2\gamma^{-1}\left( \mathcal{A}_3 + \xi \mathcal{B}_3\right)  \label{scalarforcex} \\
f_{(e)}^z & = -e^2\gamma\left(\mathcal{C}_3-2v\mathcal{A}_1  -2 \gamma^{-2}\mathcal{A}_2-v\mathcal{B}_1+(1+v^2)\mathcal{B}_2\right) \label{emforcez} \\
f_{(e)}^x & = -e^2\gamma\left(\mathcal{C}_2+v\mathcal{C}_1-2\gamma^{-2}\mathcal{A}_3 + (1+v^2)\mathcal{B}_3\right) \label{emforcex} \\
f_{(m)}^z &= -2m^2 \gamma\left(\gamma^{-2}\mathcal{A}_2-2\mathcal{C}_3+v\mathcal{A}_1 +2v(2+\gamma^2v^2)\mathcal{B}_1 -2v^2\mathcal{B}_2\right) \label{gravityforcez} \\
f_{(m)}^x & = -2m^2 \gamma \left(\gamma^{-2}\mathcal{A}_3 -2\mathcal{C}_2-2v\mathcal{C}_1 -2v^2\mathcal{B}_3\right). \label{gravityforcex} 
\end{align}
The definitions and results for the integrals are\footnote{Note that the form of the self-force in terms of the master integrals holds for any weakly curved spacetime, but the expressions we calculate for the integrals are valid only at large distances from a Newtonian star.}
\begin{align}
\mathcal{A}_1 & = \int_{-\infty}^{t^-} \frac{d}{dt}A_{,tt'} dt' = \frac{M v^2 \left(r^2 \left(1-3 v^2\right)+4 r v z+ z^2\left(v^2-3\right)\right)}{2 r (r-v z)^4} \label{calA1} \\
\mathcal{A}_2 & = \int_{-\infty}^{t^-} A_{,tt'z} dt' 
   = \frac{M v \left(r-3 v z \right) \left(r^2(1-3 v^2) +4 r z v + z^2 (v^2-3)\right)}{6 r \left(r-v z\right)^5} \\
\mathcal{A}_3 & = \int_{-\infty}^{t^-} A_{,tt'x} dt'= \frac{M v b \left(2 r^2 v^3-r z \left(1+3 v^2 \right)-\left(v^2-3\right) v z^2\right)}{2 r(r-v z)^5}, \label{calA3}
\end{align}
and
\begin{align}
\mathcal{B}_1 &= \int_{-\infty}^{t^-} \frac{d}{dt}B dt' = \frac{M v \left(r^2 v-2 r z+v z^2\right)}{r^3 (r-v z)^2} \label{calB1} \\
 \mathcal{B}_2 & = \int_{-\infty}^{t^-} B_{,z} dt' = -\frac{M z \left(r^2 \left(v^2+1\right)-3 r v z+v^2 z^2\right)}{r^3 (r-v z)^3}
 \label{calB2} \\
 \mathcal{B}_3 & =
\int_{-\infty}^{t^-} B_{,x} dt' = -\frac{M b \left(r^2 \left(v^2+1\right)-3 r v z+v^2 z^2\right)}{r^3 (r-v z)^3}, \label{calB3}
\end{align}
and
\begin{align}
\mathcal{C}_1 & = \int_{-\infty}^{t^-} \frac{d}{dt}(A_{,zx'}-A_{,xz'}) dt' = \frac{M v^2 b \left(2 r^2 v-3 r z+v z^2\right)}{r^3 (r-v z)^3} \label{calC1} \\
\mathcal{C}_2 & = \int_{-\infty}^{t^-} \frac{d}{dt}(A_{,xt'}-A_{,tx'}) dt' = \frac{M v b \left(2 r^2 v-3 r z+v z^2\right)}{r^3 (r-v z)^3} \label{calC2} \\
\mathcal{C}_3 & = \int_{-\infty}^{t^-} \frac{d}{dt}(A_{,zt'}-A_{,tz'}) dt' = \frac{M v \left(r^3 \left(1-2 v^2\right)+3 r^2 v z-3 r z^2+v z^3\right)}{r^3 (r-v z)^3}. \label{calC3}
\end{align}
In formulating these integrals, we have changed variables from $\tau'$ to $t'$, with the notation $t^-$ indicating that the integration is to be stopped at $t^-+\epsilon$ for $\epsilon>0$, after which the $\epsilon \to 0$ limit is to be taken.  The unprimed total derivative is defined to be $d/dt = \partial_t + v \partial_z$, or equivalently ordinary differentiation after evaluation on the worldline.  In these expressions, it is implicit that both the primed and unprimed points are to be evaluated on the worldline $(x=x'=b, y=y'=0, z=v t, z'=v t')$ after derivatives are taken.  The displayed results of these integrals are likewise to be evaluated at the present position of the particle, i.e., $z=vt$ and $r=\sqrt{b^2+v^2 t^2}$.  In this sense, the nine integrals are functions of $t$ alone.

Our results for the electromagnetic self-force agree with  Ref.~\cite{westpfahl1980relativistic}.  Our results for the gravitational self-force are new, although analogous statements undoubtedly exist inside Westpfahl's 2PM calculation \cite{Westpfahl:1985tsl}.  Our results for the scalar self-force are new.

\subsection{Method of computation}

  We now describe our method of computation for the integrals \eqref{calA1}-\eqref{calC3}.  As explained in Sec.~\eqref{sec:green} above, the Green functions contain terms supported on $T=R$ (the direct portion, supported on the past light cone), terms supported on $T=r-r'$ (the part of the tail supported on the secondary light cone), and terms supported on $T < r-r'$ (the remainder of the tail).  Given a spacetime point $(t,x,y,z)$, the intersection of its past light cone with the worldline defines a retarded time $t_1$, while the intersection of its secondary light cone defines a ``doubly retarded'' time $t_2$.  Since the direct portion of the Green function by construction does not contribute to the self-force, the important time is the doubly-retarded one,
\begin{align}\label{t2}
    t_2 & = \gamma^2 \left(t-r-\sqrt{b^2 \gamma^{-2}+v^2(t-r)^2} \right).
\end{align}
We also define doubly-retarded versions of some associated quantities,
\begin{align}
    r_2 & = \sqrt{b^2 + v^2 t_2^2} \\
    T_2 & = t-t_2 \\
    R_2 & = \sqrt{(x-b)^2+y^2+(z-v t_2)^2}.
\end{align}
The delta functions $\delta(T-r-r')$ appearing in the Green function may be changed to the integration variable $t'$ via
\begin{align}\label{delta}
    \delta(t-t'-r-\sqrt{b^2+v^2 t'^2}) = \frac{\delta(t'-t_2)}{1+v^2 \frac{t_2}{r_2}} \qquad (\textrm{if } t'<t),
\end{align}
where the restriction to $t'<t$ has allowed us to drop a second, unphysical root that does not contribute to the integral involving only $t'<t$.

Using Eqs.~\eqref{t2} and \eqref{delta}, the evaluation of the integrals is straightforward (especially for computer algebra software), but produces unduly complicated expressions.  In order to obtain the simple forms given in Eqs.~\eqref{calA1}-\eqref{calC3}, we have made judicious choices (described below) of the order of differentiation, evaluation on the worldline, and integration, and also employed a number intermediate formulas holding after evaluation on the worldline.  These formulas are
\begin{align}
    t_2 & \overset{z}{=} \gamma^2\left( (1+v^2)t - 2 r \right), \label{useful1} \\
r_2 & \overset{z}{=} \gamma^2( (1+v^2)r-2 v^2 t),
\label{useful2} \\
r_2 +v^2 t_2 & \overset{z}{=} r-v^2 t, \label{useful3} \\
\partial_r t_2 & \overset{z}{=} \frac{-r_2}{r-v^2 t}, \label{useful4} \\
\partial_r r_2 & \overset{z}{=} \frac{-v^2t_2}{r-v^2 t}, \label{useful5} 
\end{align}
as well as
\begin{align}
    R &\overset{z}{=} v T \label{useful6} \\ 
    T_2 & \overset{z}{=} 2 \gamma^2(r-v^2 t) \label{useful7}  \\
    T_2^2 -R_2^2 & \overset{z}{=} 4 \gamma^2(r-v^2 t)^2 \label{useful8} \\
    \partial_x(T_2^2 -R_2^2) &\overset{z}{=} \frac{-4 \gamma^2 b}{r} ((1+v^2)r-2v^2 t) \label{useful9} \\
    \partial_z(T_2^2 -R_2^2) &\overset{z}{=} \frac{4 \gamma^2 v}{r} (r^2-rt(1+2v^2)+2v^2 t^2),\label{useful10} 
\end{align}
where the symbol $\overset{z}{=}$ indicates that the equality holds when evaluated on the worldline ($x=b,y=0,z=vt$) after differentiation.   The partial derivative with respect to $r$ is used only for $t_2$ and $r_2$, in which case it means that $t$ is held fixed.

We now illustrate the method of computation with a few examples, starting with $\mathcal{B}_1$.  Plugging Eq.~\eqref{B2} into the integral in \eqref{calB1}, we have
\begin{align}
    \mathcal{B}_1 = \left.\left( \int_{-\infty}^t  \frac{d}{d t} \left.\frac{M\delta(t-t'-r-r') }{r r'}\right|_{r'=\sqrt{b^2+v^2 t'^2}} dt' \right)\right|_{x=b,\ y=0,\ z=vt},
\end{align}
where we remind the reader that $d/dt = \partial_t + v \partial_z$.  Since the integrand vanishes in a neighborhood of $t'=t$ (having support only at $t'=t-r-r'$), we have replaced the original upper limit $t^-$ with $t$.  We can also pull the time derivative out of the integral for the same reason.  Furthermore, for this total derivative we may evaluate on the worldline ``early,'' i.e.
\begin{align}\label{early}
    \mathcal{B}_1 =  \frac{d}{d t} \int_{-\infty}^t  \left.\frac{M\delta(t-t'-r-r') }{r r'}\right|_{\tiny \mathstack{r=\sqrt{b^2+v^2 t^2}}{r'=\sqrt{b^2+v^2 t'^2}}} dt'.
\end{align}
Using the delta function change of variables \eqref{delta}, the integral is calculated to be
\begin{align}
    \mathcal{B}_1 & = \frac{d}{d t} \int_{-\infty}^t \left. \frac{M}{rr'} \frac{\delta(t'-t_2) }{1+v^2 \frac{t'}{r'}}\right|_{\tiny \mathstack{r=\sqrt{b^2+v^2 t^2}}{r'=\sqrt{b^2+v^2 t'^2}}} dt' \\
    & = \frac{d}{dt}\left( \left. \frac{M}{r} \frac{1 }{r_2 +v^2 t_2}\right|_{r=\sqrt{b^2+v^2 t^2}} \right) \\
    & = \frac{d}{dt}\left( \left. \frac{M}{r(r-v^2 t)}\right|_{r=\sqrt{b^2 + v^2 t^2}}\right) \\
    & = -\left.\frac{M v^2(b^2+2t(t v^2+r))}{r^3(r-v^2 t)^2}\right|_{r=\sqrt{b^2 + v^2 t^2}}, \label{moo}
\end{align}
where we use \eqref{useful3} in the third step.  Eq.~\eqref{moo} reproduces the claimed result \eqref{calB1} after use of $z=v t$.  

The integrals for $\mathcal{B}_2$ and $\mathcal{B}_3$ proceed similarly, except that the evaluation on the worldline must now be done after the unprimed derivative is taken.  After moving the derivative outside the integral and performing the integral using Eq.~\eqref{delta}, we are left with
\begin{align}
    \mathcal{B}_2 = \left. \frac{\partial}{\partial z} \left( \frac{M}{r} \frac{1}{r_2+v^2 t_2} \right)\right|_{x=b, \ y=0, \ z=vt}.
\end{align}
The expression for $\mathcal{B}_3$ identical except the derivative is $\partial/\partial x$.  These derivatives may be evaluated and expressed back in terms of $r$ and $t$ using Eqs.~\eqref{useful1}-\eqref{useful5}, resulting in Eqs.~\eqref{calB2} and \eqref{calB3} for $\mathcal{B}_2$ and $\mathcal{B}_3$.

We next turn to the $\mathcal{C}$ integrals.  These are composed of derivatives of the biscalar $A$, which in general have (1) delta functions at $T=R$, (2) delta functions at $T=r+r'$, and (3) smooth functional dependence away from these special points.  For the special combination of derivatives appearing in the $\mathcal{C}$ integrals, however, the smooth parts cancel out, leaving only the delta functions.  The delta functions at $T=R$ are part of the direct portion of the Green function and do not contribute to the integration range $t'<t$.  What remains is supported purely at $T=r-r'$,
\begin{align}
    A_{,zx'}-A_{,xz'} & = \frac{2M(zx'-xz')}{rr'(r+r'-R)(r+r'+R)} \delta(T-r-r'), \qquad (\textrm{if } T \neq R) \\
    A_{,xt'}-A_{,tx'} & = \frac{2M}{T^2-R^2}\left( \frac{x}{r} + \frac{x'}{r'} \right) \delta(T-r-r'), \qquad \quad \qquad (\textrm{if } T \neq R) \\
     A_{,zt'}-A_{,tz'} & = \frac{2M}{T^2-R^2}\left( \frac{z}{r} + \frac{z'}{r'} \right) \delta(T-r-r'), \quad \qquad \qquad (\textrm{if } T \neq R).
\end{align}
The method of evaluation of the $\mathcal{C}$ integrals parallels that of $\mathcal{B}_1$, where one can first evaluate on the worldline, then perform the integral, and finally take the total derivative.  The results may be simplified using Eqs.~\eqref{useful1}-\eqref{useful10}.

Finally, we turn to the $\mathcal{A}$ integrals, which are constructed from derivatives of $A_{tt'}$.  Over the range of integration $t<t'$ we may again drop the ``direct'' terms supported on $T=R$, which leaves
\begin{align}
    A_{,tt'} = \frac{4MT}{(R^2-T^2)^2} \Theta(T-r-r') + \frac{2M}{R^2-T^2}\delta(T-r-r') \qquad (\textrm{if } T \neq R).
\end{align}
Evidently, the $\mathcal{A}$ integrals involve terms  proportional to $\Theta(T-r-r')$ in addition to the delta functions we have already encountered.  The Heaviside integrands turn out to be very simple once evaluated on the worldline---for $\mathcal{A}_2$ and $\mathcal{A}_3$ this means the relevant derivative must be taken \textit{before} integration---and are easily expressed in terms of anti-derivatives evaluated at the doubly-retarded time $t_2$.  The delta-function terms can be treated as before, and full expressions can be simplified using the formulas \eqref{useful1}-\eqref{useful10}.

\subsection{Limits}

Eqs.~\eqref{scalarforcez}-\eqref{calC3} provide expressions for the scalar, electromagnetic, and gravitational self-forces as functions of time $t$, expressed as a ratio of polynomials in $v$, $b$, $r=\sqrt{b^2+v^2 t^2}$, and $z=vt$.  Although relatively compact expressions can be obtained in the electromagnetic case, we find no additional insight from writing out these polynomials.  However, it is helpful to examine the limits of low and high velocity.  

For the low-velocity limit, we consider $v$ to be small but allow $vt$ to have any size, since the range of $t$ is unbounded.  Thus we expand Eqs.~\eqref{scalarforcez}-\eqref{gravityforcex} in $v$ at fixed $r$ and $z$.  Keeping through $O(v)$, we find that the results can be repackaged in vector notation as
\begin{align}
    \bm{f}_{(q)} & = q^2 \frac{2M}{r^3} \bm{\hat{r}} + \frac{1}{3} q^2 \frac{d \bm{g}}{dt} + O(v^2), \label{fqlow} \\
    \bm{f}_{(e)} & = e^2 \frac{M}{r^3} \bm{\hat{r}} + \frac{2}{3} q^2 \frac{d \bm{g}}{dt} + O(v^2), \label{felow} \\
    \bm{f}_{(m)} & = - \frac{11}{3} m^2 \frac{d \bm{g}}{dt} + O(v^2),\label{fmlow}
\end{align}
where $\bm{g}=-M/r^2 \bm{\hat{r}}$ is the Newtonian gravitational acceleration and $\bm{\hat{r}}$ is the radial unit vector.  This reproduces the low-velocity results derived in Refs.~\cite{DeWitt:1960fc,Pfenning:2000zf}, where we have made the additional assumption of straight-line motion.  As observed in these references, the scalar and electromagnetic forces each consist of a dissipative piece equal to the self-force on a particle moving in flat spacetime subject to a Newtonian gravitational force (terms proportional to $d\bm{g}/dt$), together with a conservative force that must be attributed purely to the curvature of spacetime (terms proportional to $M/r^3$).  The gravitational force contains a dissipative-type term with a ``wrong sign'' suggestive of radiation \textit{anti}-damping instead of damping.  While perhaps surprising, this should not be alarming since (1) the gravitational self-force is gauge-dependent, with the particle position not directly observable and (2) there are additional, matter-mediated forces that must be taken into account in this problem (Sec.~\ref{sec:matter-mediated} below).  These issues were first discussed in Ref.~\cite{Pfenning:2000zf} in the case of bound motion.

For the large-velocity limit, we begin by substituting $v=\sqrt{1-\gamma^{-2}}$ in Eqs.~\eqref{scalarforcez}-\eqref{gravityforcex} and expanding for large $\gamma$.  We find that the coefficients in this large-$\gamma$ series blow up at $r=z$, a behavior that originates from the denominators in Eqs.~\eqref{calA1}-\eqref{calC3}.  This divergence at $t \to -\infty$ signals the need for a separate expansion adapted to the distant past, which can be matched to the usual expansion to provide a uniformly valid approximation.  The need for a second expansion is physically natural since the time to bounce a light signal off the star and return to the particle diverges in the ultrarelativistic limit.

\section{Matter-mediated Force}\label{sec:matter-mediated}

The gravitational self-force provides an $O(mM/b^2)$ correction to the acceleration of the particle, which may be interpreted as the action of the particle's own gravitational field on its motion.  A second physical effect acts at this same perturbative order: the particle's gravitational field accelerates the star at $O(m/b)$, and the new motion of the star changes the acceleration of the particle at $O(mM/b^2)$.  Before tackling the calculation, it is helpful to review the formal origin of this additional force, which arises for any spacetime containing matter \cite{Pfenning:2000zf}.

The derivation \cite{Mino:1996nk,Quinn:1996am,Gralla:2008fg,Pound:2009sm} of the gravitational self-force assumes that the background spacetime is a vacuum solution of Einstein's equation.  However, the key assumptions are local to the particle and will still hold if the particle is restricted to a  \textit{vacuum region} of an otherwise non-vacuum spacetime.  The main difference is that the (far-zone) metric perturbation $h_{\mu \nu}$ will have a new source term $\delta T_{\mu \nu}^{\rm star}$ representing the perturbation to the background stress-energy that is induced by the presence of the particle, in addition to the point-particle stress-energy $T_{\mu \nu}^{\rm particle}$ of the particle itself.  (For clarity we refer to the perturbed matter stress-energy as that of a star, although our comments apply more generally.)  The metric perturbation similarly gains an extra term,
\begin{align}
    h_{\mu \nu} = h_{\mu \nu}^{\rm particle} + h_{\mu \nu}^{\rm star},
\end{align}
with
\begin{align}
    h_{\mu \nu}^{\rm particle} & = \int G_{\mu \nu}{}^{\mu' \nu'} T_{\mu' \nu'}^{\rm particle} \sqrt{-g} d^4 x, \label{hself} \\
    h_{\mu \nu}^{\rm star} & = \int G_{\mu \nu}{}^{\mu' \nu'} \delta T_{\mu' \nu'}^{\rm star} \sqrt{-g} d^4 x, \label{hstar}
\end{align}
where $G^{\mu \nu}{}_{\mu' \nu'}$ is the gravitational Green function defined in Sec.~\ref{sec:green} above.  We may regard $h_{\mu \nu}^{\rm particle}$ as the metric perturbation generated directly by the particle (i.e., the self-field, which diverges at the particle) and $h_{\mu \nu}^{\rm star}$ as the metric perturbation due to the shift in the star's stress-energy induced by the presence of the particle (which is smooth at the particle).  Both perturbations are proportional to the particle mass $m$; every use of $h_{\mu \nu}$ in this paper represents a term linear in $m$.  The two terms propagate through the derivation, giving two corresponding terms in the final force on the particle,
\begin{align}
    f^\mu = f_{(m)}^\mu +  f^\mu_{\rm mm}.
\end{align}
The first term is standard expression \eqref{fm} for the self-force, while the second is an additional ``matter-mediated'' force \cite{Pfenning:2000zf},
\begin{align}\label{fmm}
    f^\mu_{\rm mm} =-\frac{1}{2} m \left( g^{\mu \nu} + u^\mu u^\nu \right) \left( 2 \nabla_\beta h_{\alpha \nu}^{\rm star} - \nabla_\nu h_{\alpha \beta}^{\rm star}  \right)  u^\alpha u^\beta.
\end{align}
The matter-mediated force takes the form of a perturbed geodesic equation, i.e., ignoring the self-force, the particle would move on a geodesic of $g_{\mu \nu}+h_{\mu \nu}^{\rm star}$.

This elegant presentation of the matter-mediated force belies a severe practical difficulty: the perturbed star stress-energy $\delta T^{\mu' \nu'}_{\rm star}$ will in general be required to satisfy additional equations that involve the metric perturbation $h^{\rm star}_{\mu \nu}$ (and $h^{\rm particle}_{\mu \nu}$), making Eq.~\eqref{hstar} of little use in actually computing $h^{\rm star}_{\mu \nu}$.  Following Ref.~\cite{Pfenning:2000zf}, we will be able to circumvent this difficulty by taking advantage of the series expansion in $M$ and modeling the star as a delta function to leading order.  We are confident in the delta-function assumption because (1) no infinities arise in the subsequent calculation and (2) the more careful arguments given above for the computation of the Green function are equivalent to the assumption of a delta-function star.  Still, we emphasize that this choice has not been justified with the same rigor as analogous claims made in Sec.~\ref{sec:green} above.

In order to obtain the matter-mediated force \eqref{fmm} at $O(m^2M)$ as desired, we will need the star's perturbation $h^{\rm star}_{\mu \nu}$ at order $O(mM)$.  To determine this perturbation from Eq.~\eqref{hstar}, we will need the stress-energy of the star at $O(mM)$.  In a given spacetime $\hat{g}_{\mu \nu}$ with timelike coordinate $t$, a point particle stress-energy tensor takes the form
\begin{align}\label{Tpp}
    T^{\mu \nu}_{\rm p.p.} =  M U^\mu U^\nu \frac{\delta(\bm{x}-\bm{Z}(t))}{U^t\sqrt{-\hat{g}}}, \qquad U^\mu \hat{\nabla}_\mu U^\nu = 0,
\end{align}
where the second equation follows from conservation of stress-energy and indicates that $\bm{Z}(t)$ (four-velocity $U^\mu$) is a geodesic of the spacetime.   We will assume that this form holds for our star at first order in $M$.  Consistent to this order, we may drop all $M$-dependent terms in $\bm{Z}$, $U^\mu$, and $\hat{g}$.  We will also drop terms of $O(m^2)$, since these are neglected everywhere in our calculation.  Thus we use that
\begin{align}\label{ghat}
    \hat{g}_{\mu \nu} = \eta_{\mu \nu} + h^{\rm pf}_{\mu \nu} + O(M) + O(m^2),
\end{align}
where $h^{\rm pf}_{\mu \nu}$ (pf for ``particle flat'') is the metric perturbation due to the particle at $O(M^0 m^1)$, i.e., the leading piece of $h^{\rm particle}$.  This is just the Lorenz-gauge linearized field of a point particle moving on a straight line in flat spacetime, which may be obtained by translating and boosting the linearized Schwarzschild metric in isotropic coordinates (or, alternatively, constructed using the $M=0$ limit of the Green function discussed in Sec.~\ref{sec:green} with a point particle source).  For our choices \eqref{z} and \eqref{u} for the particle worldline, the non-zero components are 
\begin{align}
    h^{\rm pf}_{00} & = h^{\rm pf}_{zz} = \gamma^2(1+v^2)\frac{2m}{r_p} \\ 
    h^{\rm pf}_{0z} & = -2\gamma^2 v \frac{2m}{r_p} \\
    h^{\rm pf}_{xx} & = h^{\rm pf}_{yy} = \frac{2m}{r_p},
\end{align}
where
\begin{align}
    r_p = \sqrt{(x-b)^2+y^2+\gamma^2(z- v t)^2}.
\end{align}
Assuming that the point particle form \eqref{Tpp} holds for the star's stress-energy to $O(M)$ then gives 
\begin{align}
    T^{\rm star}_{00} & = M \delta(\bm{x}) - M Z^i(t) \partial_i \delta(\bm{x}) +  \frac{mM\gamma^2(3v^2-1)}{\sqrt{b^2+\gamma^2 v^2 t^2}}\delta(\bm{x}) + O(m^2 M) + O(M^2), \label{T00star} \\
    T^{\rm star}_{0i} & = - M \frac{d Z^i}{dt} \delta(\bm{x}) +O(m^2 M) + O(M^2), \label{Tiostar} \\
    T^{\rm star}_{ij} & = O(m^2 M) + O(M^2), \label{Tijstar}
\end{align}
where we have used the fact that $Z^i=O(m)$ to expand the delta function.

The motion of the star $Z^i(t)$ is determined by the geodesic equation in the spacetime $\hat{g}_{\mu \nu}$ with $Z^\mu=(t,0,0,0)$ to leading order, which takes the form
\begin{align}\label{geod}
    \frac{d^2 Z^i}{dt^2} = \frac{1}{2} \partial_i h^{\rm pf}_{00} - \partial_0 h^{\rm pf}_{0i} + O(M m) + O(m^2).
\end{align}
 Integrating Eq.~\eqref{geod} once, we find for the velocity $V^i=dZ^i/dt$ that
\begin{align}
    V^z(t) & = \frac{m\gamma(3v^2-1)}{v \sqrt{b^2\gamma^{-2}+ v^2 t^2}} \label{Vz} \\
    V^x(t) & = \frac{ m\gamma(1+v^2)}{bv}\left(1+\frac{vt}{\sqrt{b^2\gamma^{-2}+v^2 t^2}} \right),\label{Vx}
\end{align}
where we suppress the $O(Mm)$ and $O(m^2)$ error terms.  We have fixed the integration constants by demanding that the star velocity vanish at early times.  Integrating again, we find for the position $Z^i=(X,Y,Z)$ that
\begin{align}
    Z(t) & = \frac{m \gamma(3v^2-1)}{v^2}\textrm{arctanh} \frac{v t}{\sqrt{b^2\gamma^{-2}+ v^2 t^2}} \label{Zz} \\
   X(t) & = \frac{m\gamma(1+v^2)}{bv^2} \left( vt + \sqrt{b^2\gamma^{-2} + v^2 t^2} \right). \label{Zx}
\end{align}
For the $X$ position, we have chosen the integration constant such that $X(t\to -\infty)=0$.  The analogous choice for the $Z$ position is not possible since $Z(t)$ diverges logarithmically and early (and late) times.  In the low-velocity limit of ordinary Newtonian dynamics, this divergence can be attributed to the $1/r^2$ force, which gives a log when twice integrated.  Here we have chosen the integration constant so that $Z=0$ at $t=0$.

The metric perturbation $h_{\mu \nu}^{\rm star}$ of the star is given by Eq.~\eqref{hstar}.  To determine the $O(mM)$ metric perturbation we will need only the $O(M^0)$ part of the Greeen function, i.e., the flat spacetime Green function.  From the fundamental equation \eqref{Ggravity}, the trace-reversed Green function is simply related to the scalar Green function $G_{\rm flat}$ \eqref{Gflat} as
\begin{align}
    \bar{G}^{\alpha \beta}{}_{\alpha' \beta'} = \delta^{(\alpha}{}_{\alpha'}\delta^{\beta)}{}_{\beta'} \frac{\delta(t-t'-|\bm{x}-\bm{x}'|)}{|\bm{x}-\bm{x}'|} +O(M),
\end{align}
meaning that we can determine $h^{\rm star}_{\mu \nu}$ to the relevant order $O(Mm)$ by
\begin{align}
    h^{\rm star}_{\mu \nu} = \bar{h}^{\rm star}_{\mu \nu}-\frac{1}{2} \eta_{\mu \nu} \eta^{\alpha \beta} \bar{h}^{\rm star}_{\alpha\beta}, \qquad  \bar{h}^{\rm star}_{\mu \nu} = 4 \int \frac{\delta T^{\rm star}_{\mu \nu} (t-|\bm{x}-\bm{x}'|,\bm{x}')}{|\bm{x}-\bm{x'}|}d^3 x'.
\end{align}

The star's perturbed stress-energy tensor $\delta T_{\mu \nu}^{\rm star}$ is identified as the $O(m)$ piece of Eqs.~\eqref{T00star}-\eqref{Tijstar} (recalling that $Z^i=O(m)$), and performing the integral gives
\begin{align}
    h^{\rm star}_{00} & = \frac{2M}{r} \left[ \left( \frac{ Z_i|_{t-r}}{r} + V_i|_{t-r} \right) \frac{x^i}{r} + \frac{m \gamma (3v^2-1)}{\sqrt{b^2\gamma^{-2}+ v^2(t-r)^2}} \right] \label{hstar00} \\
    h^{\rm star}_{0i} & = -\frac{4M V^i|_{t-r}}{r}  \\
    h^{\rm star}_{ij} & = h^{\rm star}_{00} \delta_{ij}, \label{hstarij} 
\end{align}
where the notation $Z_i|_{t-r}$ means to evaluate the function $Z_i(t)$ at $t-r$, i.e., $Z_i|_{t-r}=Z_i(t-r)$ (and likewise for $V^i$).  These functions are given in Eq.~\eqref{Vz}-\eqref{Zx} above.

The matter-mediated force is now given by Eq.~\eqref{fmm}.  As $h^{\rm star}_{00}$ is already $O(M)$, the $O(M^0)$ metric and particle four-velocity may be used, giving
\begin{align}
f^z_{\rm mm} & = \frac{m \gamma^2}{2} \left( (1+v^2) h^{\rm star}_{00,z} + 2 v h^{\rm star}_{0z,z} - \gamma^2 v(3-v^2) \frac{d}{dt}h^{\rm star}_{00} - 2 \gamma^2 \frac{d}{dt}h^{\rm star}_{0z} \right) \label{fmmz} \\
    f^x_{\rm mm} & = \frac{m \gamma^2}{2}\left( (1+v^2)h^{\rm star}_{00,x}+2 v h^{\rm star}_{0z,x} - 2 \frac{d}{dt} h^{\rm star}_{0x} \right), \label{fmmx}
\end{align}
where the terms are to be evaluated on the particle worldline $(x=b,y=0,z=vt)$.  In writing this expression, we have used Eq.~\eqref{hstarij} to eliminate $h^{\rm star}_{ij}$.  

To simplify the expressions \eqref{fmmz} and \eqref{fmmx} for the matter-mediated force, it is helpful to note that the  square root in \eqref{hstar00} is related to the doubly-retarded time $t_2$ defined in \eqref{t2} by
\begin{align}
    \sqrt{b^2\gamma^{-2} + v^2(t-r)^2} = t - r - \gamma^{-2} t_2.
\end{align}
This square root also arises in the expressions \eqref{Vz}-\eqref{Zx} for $V_i$ and $Z_i$ after sending $t\to t-r$, as appears in  $h^{\rm star}_{\mu \nu}$ via Eqs.~\eqref{hstar00}-\eqref{hstarij}.  After eliminating all such square roots in favor of $t_2$, one may use Eqs.~\eqref{useful1}-\eqref{useful5} to simplify the expressions when evaluated on the worldline.  Doing so, we find
\begin{align}
    f^z_{\rm mm} & = \frac{\gamma^5 m^2 M}{v^2 r^5(r- vz)^3} \mathcal{F}_z(r,v,z) \label{fmmzresult} \\
    f^x_{\rm mm} & = \frac{\gamma^3 m^2 M}{v^2 r^5(r- vz)^3} \frac{\mathcal{F}_x(r,v,z)}{b},\label{fmmxresult}
\end{align}
where

\begin{align}
\mathcal{F}_z &= r^5 v \left(2
   v^8-8 v^6+v^5+9 v^4-2 v^3+2 v^2-3 v-1\right) \nonumber \\   &-r^4 z\left(4 v^8+3 v^7-5 v^6-6 v^5+37 v^4-9 v^3-15 v^2+3\right)  \nonumber \\
   & +3 r^3 z^2\left(v^8+5 v^7-2 v^6+18 v^5-15 v^3+2 v^2+4 v-1\right) \nonumber \\ 
&-r^2  z^3 v \left(v^8+5 v^7-2
   v^6+48 v^5+24 v^4-67 v^3+18 v^2+18 v-9\right) \nonumber \\ 
   &+3 r  z^4 v^2 \left(9 v^4-12 v^3+6 v^2+4 v-3\right) \nonumber\\
   & - 3  z^5 v^3 \left(-3 v^5+3
   v^4-2 v^3+2 v^2+v-1\right)\nonumber \\
   &- \left(r^2-3 z^2\right) (v z-r)^3 \left(1-3 v^2\right)^2 \textrm{arctanh}[(r v - z)/(r - v z)]
\end{align} 
and
\begin{align}
\mathcal{F}_x &= -2 r^6 (1 + v^2) (1 - 4 v^4 + v^6) \nonumber \\ &-
 r^5 z (2 - 9 v + 8 v^2 - 12 v^3 + 6 v^4 + 39 v^5 - 6 v^7) \nonumber \\ &+ 
   r^4 z^2 (3 + 6 v - 15 v^2 + 24 v^3 - 23 v^4 + 18 v^5 + 11 v^6) \nonumber \\ &- 
   r^3 z^3 (-3 + 12 v - 5 v^3 + 21 v^4 - 62 v^5 + 18 v^6 + 3 v^7) \nonumber \\ &-
   r^2 z^4 v (9 - 18 v + 16 v^2 - 3 v^3 + v^4 + 32 v^5 - 6 v^6 + v^7) \nonumber \\ 
      &+ 3 r  z^5 v^2 (1 + v^2) (3 - 4 v + 3 v^2) \nonumber \\
      & + 3 z^6 (-1 + v) v^3 (1 + v^2)^2 \nonumber \\ 
      &- 3 b^2  z (-r + v z)^3 (1 + v^2) (-1 + 3 v^2) \textrm{arctanh} [(r v - z)/(r - v z)].
\end{align}
This completes the calculation of the matter-mediated force.

Note that the matter-mediated force arises only because our background spacetime contains matter.  If we had repeated our calculation using a black hole background instead of a star background, there would be no matter-mediated force at all, and the entire $O(mM)$ dynamics would be given by the gravitational self-force.  Since we expect the binary dynamics to be independent of the body compositions at this order of approximation, the natural conclusion is that the Lorenz-gauge gravitational self-force must depend in detail on the central body composition, even in the limit where the particle is very distant compared to the typical size of the central body.  This phenomenon has been seen previously for nonminimally coupled scalar fields, but not for minimally coupled scalar fields or electromagnetic fields \cite{Pfenning:2000zf,Drivas:2010wz,Isoyama:2012in,Chu:2019vmh}.  We note, however, that the issue of the back-reaction of black hole motion on the particle is closely tied up with the question of the choice of gauge \cite{Detweiler:2003ci,Barack:2019agd}.

\section{Motion of the particle}\label{sec:motion}

In any spacetime, the equations of motion for a mass $m$ moving under a four-force $f^\mu$ are
\begin{align}\label{eom}
   m \frac{d^2 x^\mu}{d\tau^2} = - m\Gamma^\mu_{\alpha \beta}\frac{dx^\alpha}{d\tau} \frac{dx^\beta}{d\tau} + f^\mu.
\end{align}
We will integrate these equations in the weakly curved spacetime \eqref{spacetime} with background particle trajectory \eqref{z}, where the force is either the scalar self-force, the electromagnetic self-force, or the sum of the gravitational self-force and the matter-mediated force.  In this section we work consistently to $O(M)$, dropping all terms of order $O(M^2)$ or higher.  Expanding \eqref{eom} gives
\begin{align}
    \frac{ d^2t}{d\tau^2} & = - \gamma^2 \frac{2 M z}{r^3} v + \frac{v f^z}{m}  \label{d2t} \\
    \frac{ d^2z}{d\tau^2} & = - \frac{M z}{r^3} + \frac{f^z}{m} \label{d2z} \\
    \frac{ d^2x}{d\tau^2} & = - \gamma^2 \frac{M x}{r^3} (1+v^2) + \frac{f^x}{m}, \label{d2x}
\end{align}
where the right-hand sides are to be evaluated on the background worldline ($x=b$, $z=vt$, and $r=\sqrt{b^2+v^2 t^2}$, with $t=\gamma \tau$).  In writing these equations we have used the fact that $f^0=v f^z$ to this order, as discussed above Eq.~\eqref{scalarforcez}.  In the scalar theory the rest mass $m$ is not constant (see \ref{app:mass}), but the variation begins at $O(Mq^2)$ and hence contributes an $O(M^2)$ term to Eqs.~\eqref{d2t}-\eqref{d2x}, which is neglected.  We therefore treat $m$ as a constant.

Integrating once and choosing the perturbed four-velocity to vanish in the infinite past, we find  consistent to $O(M)$ that
\begin{align}
    \frac{dt}{d\tau} & = \gamma + \frac{2 M \gamma}{r} + \frac{v}{\gamma m} \int_{-\infty}^{\gamma \tau} f^z d t', \label{dtdtau}\\
    \frac{dz}{d\tau} & = \gamma v + \frac{M}{\gamma v r} + \frac{1}{\gamma m} \int_{-\infty}^{\gamma \tau} f^z d t', \label{dzdtau} \\
    \frac{dx}{d\tau} & = -\frac{M \gamma}{b v}\left(1+v^2\right)\left(1+ \frac{z}{r} \right) + \frac{1}{\gamma m} \int_{-\infty}^{\gamma \tau} f^x d t' , \label{dxdtau}
\end{align}
where the right-hand sides are evaluated on the background worldline.  That is, we set $x=b$, $y=0$ and $z=v \gamma \tau$ in the terms without integrals, and the integrals are taken along the background worldline parameterized by $t'$, i.e., $f^z$ and $f^x$ are to be evaluated at $x=b$, $y=0$, $z=v t'$.  (The prime distinguishes this  parameter $t'$ from the $t$ appearing on the left-hand side of \eqref{dtdtau}, which represents the $t$ coordinate of the particle's perturbed motion at proper time $\tau$.)  Integrating a second time, we find
\begin{align}
    t & = \gamma \tau + \frac{2M}{v}\textrm{arctanh}\frac{z}{r} + v G^z(\gamma \tau), \label{tfull} \\
    z & = v \gamma \tau + \frac{M}{\gamma^2 v^2}\textrm{arctanh}\frac{z}{r} + G^z( \gamma \tau ), \label{zfull} \\
    x & = b - \frac{M}{bv^2}(1+v^2) (r+z) + G^x( \gamma \tau ),\label{xfull}
\end{align}

where again the background motion ($x=b$, $y=0$, $z=v\gamma\tau$) is to be inserted on the right-hand side, and we define
\begin{align}
    G^\mu(t) = \frac{1}{\gamma^2 m} \int_{-\infty}^{t} dt' \int_{-\infty}^{t'} dt'' f^\mu(t'').\label{Gdef}
\end{align}

In these integrals, the force $f^\mu$ is evaluated on the background trajectory parameterized by $t''$, i.e., with $x=b$, $y=0$, $z=v t''$.  In writing the solution this way, we have chosen the integration constants so that the background quantities retain their physical meaning as initial values.  That is, $v$ is the initial velocity (with $\gamma$ the initial Lorentz factor), and $b$ is the initial $x$ position, i.e., the impact parameter.  Note that the integration constants associated with the position integrals for $t$ and $z$ do not affect the meaning of $b$ and $v$ and have been chosen for mathematical convenience.  It is not possible to choose these so that the perturbation vanishes at early times, due to the logarithmic divergences associated with the long-range nature of the gravitational force (see discussion below Eq.~\eqref{Zx}).

Sometimes it is more convenient to use $t$ as the parameter.  From Eqs.~\eqref{tfull}-\eqref{xfull}, consistent to $O(M)$, we have
\begin{align}
    z & = v t + \frac{M(1-3v^2)}{ v^2}\textrm{arctanh}\frac{z}{r} + \gamma^{-2} G^z( t ), \label{zfullt} \\
    x & = b - \frac{M}{bv^2}(1+v^2) (r+z) + G^x( t),\label{xfullt}
\end{align}
where the right-hand-sides are now evaluated at $x=b, y=0, z=vt$.  The integrals for $G^x(t)$ and $G^z(t)$ are straightforward, if tedious, and in \ref{sec:trajectories} we present the results for $x(t)$ and $z(t)$ in the scalar, electromagnetic, and gravitational cases.  We also provide separate formulas for $G^z$ and $G^x$ in \eqref{Gzscalar}--\eqref{Gxgrav}, so that the reader can efficiently reconstruct the proper-time parameterized trajectories given in Eqs.~\eqref{tfull}-\eqref{xfull} above.  This completes the derivation of the star-frame trajectories.

\subsection{Physical Quantities}

We now use the particle trajectories to compute physical quantities defined in the star frame.  We will begin with the energy and momentum.  The four-momentum of a point particle is given by $p^\mu = m dx^\mu/d\tau$.  The initial energy-momentum $p_0^\mu$ of the particle is by definition
\begin{align}\label{pi}
    p^\mu_0 = m \gamma(1,0,0,v),
\end{align}
which also follows from the $\tau \to -\infty$ limit of Eqs.~\eqref{tfull}--\eqref{xfull}, or equivalently from the $t \to -\infty$ limit of Eqs.~\eqref{zfullt} and \eqref{xfullt}.  The final energy-momentum $p_f^\mu$ is instead calculated from the late-time limit of these equations, using the different forces $f^\mu$ as appropriate.   We will present the results as
\begin{align}
    p^\mu_{f,{\rm scalar}} &  = p^\mu_{0} + \Delta p^\mu_{(M)} + \Delta p^\mu_{(q)}  \label{pfscalar} \\
     p^\mu_{f,{\rm EM}} & = p^\mu_{0} + \Delta p^\mu_{(M)} + \Delta p^\mu_{(e)}  \label{pfem} \\
    p^\mu_{f,{\rm grav}} & = p^\mu_{0} + \Delta p^\mu_{(M)}  + \Delta p^\mu_{(m)} + \Delta p^\mu_{\rm mm} , \label{pfgrav}
\end{align}
where\footnote{In Eqs. \eqref{Deltapm}, we have corrected a typo from the previous arXiv version. We thank Oliver Long for helping find this typo.}
\begin{align}
    \Delta p^\mu_{(M)} & = \frac{M m\gamma}{bv}\Big(0,-2(1+v^2), 0, 0\Big) \label{DeltapM} \\
    \Delta p^\mu_{(q)} & = \frac{M q^2 \gamma}{b^2 v} \Big(0, \frac{\pi}{4} \left( v^2 + 4 \xi \gamma^{-2} \right),0,0\Big) \label{Deltapq} \\
    \Delta p^\mu_{(e)} & = \frac{M e^2 \gamma}{b^2 v}\Big(0, \frac{\pi}{4} \left(2+v^2\right),0,0\Big) \label{Deltape} \\
    \Delta p^\mu_{(m)} & = \frac{M m^2 \gamma v}{b^2} \Big(0, -\frac{7\pi}{4}, 0, 0 \Big) \label{Deltapm} \\
     \Delta p^\mu_{\rm mm} & = \frac{M m^2 \gamma}{b^2 v} \Big(-\frac{2 \gamma}{v} (1+v^2)^2, -\pi\left( 3-v^2 \right),0,-\frac{2 \gamma}{v^2} (1+v^2)^2 \Big). \label{Deltapmm}
\end{align}
Notice that the geodesic term \eqref{DeltapM} and the self-forces \eqref{Deltapq}, \eqref{Deltape}, and \eqref{Deltapm} only modify the $x$ component of the four-momentum, while the matter-mediated piece changes $p^t$ and $p^z$ as well.  

The momentum change is directly related to the scattering angle $\delta$, which for our initial conditions \eqref{pi} is given by
\begin{align}
    \tan \delta = -\frac{p_f^x}{p_f^z}.
\end{align}
For our small-angle scattering problem, (i.e., consistent to $O(M)$) this may be approximated by 

\begin{align}
      \delta = -\frac{p_f^x}{\gamma m v}. \label{computedelta}
\end{align}
The deflection angles then follow from Eqs.~\eqref{pfscalar}--\eqref{Deltapmm}.  The explicit expressions in the scalar, electromagnetic, and gravitational cases were displayed in Eqs.~\eqref{deltascalar_intro}, \eqref{deltaem_intro}, and \eqref{deltagravity_intro} above.  In those equations, we have also restored the explicit $O(M/b)^2$ errors as well as inserted dots ($\dots$) to indicate the presence of higher-order terms in the small parameters $q^2/(mb)$, $e^2/(mb)$, and $m/b$.   Although the scalar and electromagnetic field equations are linear, these higher-order terms will still arise at least from iterating a self-force calculation using the corrected motion.

Finally, we discuss angular momentum.  The angular momentum of a point particle about the origin of coordinates is $J^i=\epsilon^{ijk} x^j p^k$, where $x^\mu$ is the position and $p^\mu$ is the four-momentum.  The initial angular momentum $J_0^i$ of the particle is by definition
\begin{align}
   J_0 = - J^y_0 = \gamma m v b,
\end{align}
which also follows from the $\tau \to -\infty$ limit of Eqs.~\eqref{tfull}--\eqref{xfull} or the $t \to -\infty$ limit of Eqs.~\eqref{zfullt} and \eqref{xfullt}.  The final angular momentum follows from the late-time limit of these equations.  In the scalar and electromagnetic cases, we find
\begin{align}
    J_f^{\rm scalar} & = J_0 \left( 1 - \frac{2}{3} \frac{\gamma(1+v^2)}{v} \frac{M q^2}{mb^2 } \right) \\
    J_f^{\rm EM} & = J_0\left( 1 - \frac{4}{3}\frac{\gamma (1+v^2)}{v}  \frac{M e^2}{m b^2} \right).
\end{align}
In the gravitational case we find that the corresponding ``final angular momentum'' of the particle diverges logarithmically in time, indicating that the star frame is not suitable for discussing the particle angular momentum.  We will see in Sec.~\ref{sec:PM} that the difficulty disappears in the CEM frame.

\section{2PM gravitational scattering}\label{sec:PM}

In the previous section we determined the motion of the particle in the frame of the star, meaning the frame where the star has asymptotically zero velocity in the distant past.  In the scalar and electromagnetic cases, it was consistent to keep the star at rest for the purposes of computing the particle trajectories at the given order of approximation.  By contrast, in the gravitational case we required the $O(m)$ motion of the star (Eqs.~\eqref{Zx} and \eqref{Zz}) in order to determine the $O(Mm)$ motion of the particle.  However, for a complete 2PM treatment we are still missing the $O(mM)$ motion of the star as well as the second-order contributions, $O(M^2)$ and $O(m^2)$, to the motion.

Our strategy for determining these missing pieces is to determine the $O(M^2)$ motion of the particle from the geodesic equation in the Schwarzschild spacetime (\ref{sec:trajectories}) and to invoke a body exchange symmetry in order to determine the  motion of the star.  It is intuitively clear that such a symmetry should exist, since the assignment of the words ``particle'' and ``star'' to the two bodies should not affect their motion in a framework where only the respective masses $m$ and $M$ appear in the final answer.  However, our calculation has made coordinate choices which break this symmetry: our background configuration takes the star to be at rest, and our perturbation theory is correspondingly asymmetric, with gauge choices made in different ways for the star and particle.  To make the exchange symmetry manifest we will have to change coordinates to treat the bodies symmetrically. 

For clarity, let us first discuss the situation for two non-interacting point particles, which we name ``particle'' (mass $m$) and ``star'' (mass $M$) for consistency.  In this case we can define the star frame, where the star has position $(X,Y,Z)=(0,0,0)$ and the particle has position $(x,y,z)=(b,0,vt)$, and the center of energy-momentum (CEM) frame, where there is no net momentum and the center of energy is at the spatial coordinate origin.  If the star frame has Minkowski coordinates $(t,x,y,z)$ and the CEM frame has Minkowski coordinates $(\tilde{t},\tilde{x},\tilde{y},\tilde{z})$, then the frames are related by 
\begin{align}
\tilde{t} & = \frac{M+\gamma m}{\tilde{E}}t - \frac{\gamma m v}{\tilde{E}}z \label{tCM} \\
\tilde{z} & = \frac{M+\gamma m}{\tilde{E}}z - \frac{\gamma m v}{\tilde{E}}t \label{zCM} \\
\tilde{x} & = x - b\frac{m(\gamma M  + m)}{\tilde{E}^2}, \label{xCM}
\end{align}
where $\gamma=(1-v^2)^{-1/2}$ as before in this paper, and $\tilde{E}$ denotes the total energy in the CEM frame,
\begin{align}
    \tilde{E} = \sqrt{m^2+M^2+2\gamma M m}.
\end{align}
This transformation involves a boost in the $z$ direction to eliminate the net momentum, together with a translation in the $x$ direction to eliminate the non-zero center of energy.

In the CEM frame of these non-interacting bodies, swapping the bodies is equivalent to rotation by $180^\circ$.  That is, we have
\begin{align}\label{symmetry}
\tilde{Z}(\tilde{t}, M, m) =-\tilde{z}(\tilde{t}, m, M), \qquad \tilde{X}(\tilde{t}, M, m) =-\tilde{x}(\tilde{t}, m, M),
\end{align}
where $\tilde{\mathbf{Z}}=(\tilde{X},\tilde{Y},\tilde{Z})$ is the CEM-frame spatial position of the star, while $\tilde{\mathbf{z}}=(\tilde{x},\tilde{y},\tilde{z})$ is the CEM-frame spatial position of the particle.  For the gravitational scattering problem, we will \textit{define} the ``initial CEM frame'' (CEM frame for short) by the same transformation \eqref{tCM}--\eqref{xCM} applied to what we have called the initial star frame, and then \textit{impose} the swapping symmetry \eqref{symmetry} in order to determine the motion of the star.\footnote{The 1PM trajectory of the star used in the derivation of the matter-mediated force [Eqs.\eqref{Zz} and \eqref{Zx}] agrees with the 1PM trajectory determined from this method, showing that no further gauge transformations are required.}  This is logically equivalent to repeating our entire calculation with the words ``particle'' and ``star'' interchanged and finding a coordinate transformation to make the combined solution respect the exchange symmetry.

The particle trajectory in the star frame, $\mathbf{z}(t)$, is given as Eq.~\eqref{zgravfinal} together with later supporting equations.  Plugging these expressions into Eqs.~\eqref{zCM} and \eqref{xCM} gives the CEM-frame trajectory $\tilde{\mathbf{z}}(t)$, but still parameterized by star-frame time $t$.  To express in terms of the CEM-frame time $\tilde{t}$, we plug the trajectory $z(t)$ [Eq.~\eqref{zgravfinal}] into Eq.~\eqref{tCM} and solve for $t$, order by order in the PM expansion.  Accurate to 2PM, we have 
\begin{align}
t = t_0 & + \frac{mM}{m +\gamma M}\frac{\gamma^2 \left(1-3 v^2\right) }{v}\textrm{arctanh}\frac{v t_0}{ \sqrt{b^2+v^2 t_0^2}} \nonumber\\ 
&+\frac{m^2 M^2}{ (m+\gamma  M)^2}\frac{\gamma^4\left(1-3 v^2\right)^2}{\sqrt{b^2+v^2t_0^2}} \textrm{arctanh}\frac{v t_0}{ \sqrt{b^2+ v^2t_0^2}} \nonumber\\ 
& + \frac{\gamma^2 m v}{m+\gamma M}\left(z_{g2}(t_0) + z_{m}(t_0) + z_{\rm mm}(t_0)\right), \label{tcmt}
\end{align}
where we have defined
\begin{align}\label{t0}
t_0 & = \frac{\gamma \tilde{E}}{m + \gamma M} \tilde{t}.
\end{align}
This completes the derivation of the 2PM trajectories of the particle and star, expressed in the initial CEM frame. The particle trajectory $\mathbf{z}(\tilde{t})$ is given by Eqs.~\eqref{zgravfinal}, \eqref{xg1}--\eqref{zg2}, \eqref{xm}--\eqref{zmm}, \eqref{zCM}, \eqref{xCM}, and \eqref{tcmt}, and the star trajectory $\mathbf{Z}(\tilde{t})$ is then given by Eq.~\eqref{symmetry}.

\subsection{Initial conditions}

It is instructive to consider the ``initial conditions'' for the CEM-frame scattering problem, i.e., the $\tilde{t} \to -\infty$ behavior of the trajectories.  Expanding at early times, we find
\begin{align}
\tilde{\mathbf{z}} & = \left(\frac{M(M+\gamma m)}{\tilde{E}^2}b, \ 0, \frac{M \gamma}{m + \gamma M}\left[v \tilde{t} - \frac{(1-3v^2) \tilde{E}}{v^2}\log\frac{2\tilde{E}\gamma v |\tilde{t}|}{b(m+\gamma M)}\right] \right) \label{ztildeearly} +O\left(\frac{\log|\tilde{t}|}{\tilde{t}}\right) \\
\tilde{\mathbf{Z}} & = \left(-\frac{m(m+\gamma M)}{\tilde{E}^2}b, 0, \frac{-m\gamma}{M + \gamma m}\left[ v \tilde{t} - \frac{(1-3v^2) \tilde{E}}{v^2}\log\frac{2\tilde{E}\gamma v |\tilde{t}|}{b(M+\gamma m)}\right]\right) + O\left(\frac{\log|\tilde{t}|}{\tilde{t}}\right). \label{Ztildeearly}
\end{align}
 The 1PM logarithmic correction is an unavoidable consequence of the long-range nature of the gravitational force, as already discussed below Eqs.~\eqref{Zz} and \eqref{Zx}.  Notice, however, that there is no 2PM logarithmic correction.  

The final trajectories of the bodies can be analogously calculated from the $\tilde{t} \to +\infty$ limit of the trajectories derived in the previous subsection.  We will now report the initial and final values of the various conserved quantities. 

\subsection{Energy, momentum, and scattering angle}

The initial four-momentum of the particle is
\begin{align}
\tilde{p}^\mu_{0} & = m \left( \frac{m+\gamma M}{\tilde{E}}, 0, 0, \frac{\gamma M v}{\tilde{E}} \right).\label{p0CEM}
\end{align}
The final four-momentum of the particle is given as $\tilde{p}_f^\mu = \tilde{p}_0^\mu + \Delta \tilde{p}^\mu$ with 
\begin{align}
    \Delta \tilde{p}^\mu = \frac{\gamma m M}{b v} \left(0, -2 \left(v^2+1\right) -  \frac{3 \pi}{4}  \left(v^2+4\right)\frac{m+M}{b}, 0,- 2(v^2+1)^2\frac{\tilde{E}}{bv^2}\right). \label{deltapCEM}
\end{align}
The four-momentum of the star is determined by sending $M \leftrightarrow m$ and $\mathbf{x} \to -\mathbf{x}$ (i.e., flipping the sign of the spatial components of the four-momentum).  This shows that the mechanical energy is separately conserved for particle and star, and that mechanical momentum is exchanged, with the total mechanical momentum conserved.  In particular, the particle and star deflect by the same amount $\chi$ satisfying $\tan \chi=-\tilde{p}_f^x/\tilde{p}_f^z$.  Expanding this equation to 2PM order, we find
\begin{align}\label{chi-final}
\chi = \frac{\tilde{E}}{b v^2}\left( 2(v^2+1) + \frac{3\pi}{4} (v^2+4) \frac{M+m}{b} \right).
\end{align}
This result can also be derived by boosting the lab-frame deflection angle $\delta$, as we did to arrive at \eqref{chi_intro} above.

\subsection{Angular momentum}

We define the initial and final angular momentum using the special-relativistic formula evaluated at early and late times, where the particles are widely separated.  For the particle, the angular momentum formula is
\begin{align}\label{JCEM}
    \tilde{J} = -\tilde{J}^y = \tilde{p}^z \tilde{x} - \tilde{p}^x \tilde{z}.
\end{align}
Since $\tilde{x}(\tilde{t})$ and $\tilde{z}(\tilde{t})$ can diverge like $\tilde{t}$ at early and late times, in principle the $1/\tilde{t}$ corrections to $\tilde{p}^\mu$ may contribute to the early and late-time limits of Eq.~\eqref{JCEM}.  These corrections involve both the $1/\tilde{t}$ behavior of the trajectory as well as corrections due to the metric of the star.  However, we find that the $1/\tilde{t}$ terms end up canceling in the expression \eqref{JCEM} for the angular momentum at 2PM order.  In practice, this means that one may use the initial and final momenta \eqref{p0CEM} and \eqref{deltapCEM} in evaluating the angular momentum \eqref{JCEM}.

Taking the $\tilde{t}\to -\infty$ limit of Eq.~\eqref{JCEM}, we find that the initial angular momentum of the particle is given by
\begin{align}
\tilde{J}_0 & = -\tilde{J}^y_0 = b \gamma v \frac{m M}{\tilde{E}} \frac{M(M+\gamma m)}{\tilde{E}^2}.
\end{align}
From the $\tilde{t} \to +\infty$ limit, we find that the final angular momentum is given by $\tilde{J}_f = \tilde{J}_0 + \Delta \tilde{J}$ with
\begin{align}
    \Delta \tilde{J} = \frac{2 \gamma^2 M^2 m^2}{\tilde{E} b v^3} (1+v^2)\left( \frac{8}{3}v^3 -v + (1-3v^2)\textrm{arctanh}\ \! v\right).
\end{align}
The angular momentum of the star is given by sending $m \leftrightarrow M$ in these equations.  The total initial angular momentum is thus
\begin{align}
    \tilde{J}^{\rm tot}_0 = \frac{b \gamma m M v}{\tilde{E}}.
\end{align}
The particle and star each lose the same amount of angular momentum, with the fractional change in the total mechanical angular momentum given by
\begin{align}
    \frac{\Delta \tilde{J}^{\rm tot}}{\tilde{J}^{\rm tot}_0} = \frac{4 \gamma M m}{b^2 v^4}(1+v^2)\left(\frac{8}{3}v^3 - v + (1-3v^2)\textrm{arctanh}\ \! v\right).
\end{align}
This agrees with Eq.~(4.6) of Ref.~\cite{Damour:2020tta}, providing a direct check that the mechanical angular momentum lost matches the angular momentum radiated away in gravitational waves.

\subsection{Mass moment}

Information about center of energy is encoded in the time-space cross-terms of the relativistic angular momentum tensor, which we will refer to as the \textit{mass moment} $\bm{N}$.  The special-relativistic formula for the mass moment of the particle is
\begin{align}\label{NCEM}
    \tilde{\bm{N}} = \tilde{p}^0 \tilde{\bm{z}} - \tilde{\bm{p}} \tilde{t}.
\end{align}
Expanding at $\tilde{t} \to -\infty$, we find 
\begin{align}
    \tilde{N}^x & = O\left(\frac{\log|\tilde{t}|}{\tilde{t}}\right) \label{Nxearly} \\
    \tilde{N}^z & = -\gamma (1-3v^2)\frac{M m}{ v^2}\left(\log  \frac{2 v \tilde{E} |\tilde{t}|}{(m + \gamma M)b}-1 \right) + O\left(\frac{\log|\tilde{t}|}{\tilde{t}}\right) \label{Nzearly}
\end{align}
The presence of the $\log \tilde{t}$ means that the the initial particle mass moment is not well-defined.  Note also that the $-1$ in Eq.~\eqref{Nzearly} arises from a $1/\tilde{t}$ correction to the late-time momentum.  However, according to the symmetry \eqref{symmetry}, the star's early-time mass moment is given by sending $m \leftrightarrow M$ and multiplying by $-1$.  Thus these features cancel out of the total system mass moment, which has the well-defined initial value of
\begin{align}
    \tilde{N}^{x, {\rm tot}}_0 & = 0 \\
    \tilde{N}^{z, {\rm tot}}_0 & = -\gamma (1-3v^2)\frac{M m}{ v^2}\log  \frac{M + \gamma m}{m + \gamma M}.
\end{align}
That is, provided we discuss the total mass moment, we may compute from \eqref{NCEM} using the initial value \eqref{p0CEM} of the four-momentum, just as in the case of angular momentum.

The non-zero initial value of the system mass moment may suggest that we are not, after all, in the initial CEM frame.  However, we have considered only the \textit{mechanical} contribution to the mass moment, ignoring any effects of the gravitational field.  In the electromagnetic analog problem \cite{EMscoot}, a contribution from the electromagnetic field precisely cancels this mechanical portion, such that the initial mass moment is indeed zero.  The name ``initial CEM frame'' is thus fully justified in the electromagnetic case, and we will continue to use it in the gravitational case studied here.

The late-time behavior of the mass moment is precisely analogous to the early-time limit: the particle and star contributions are individually logarithmically divergent, but the total is well-defined and sensitive only to the late-time four-momentum.  The final value of the system mass moment is given by\footnote{In Eqs. \eqref{Nzf} and \eqref{Nzf2}, we have corrected an error from the previous arXiv version. We thank Hongji Wei for helping find this error.}
\begin{align}
\tilde{N}^{z,\rm{tot}}_f & = \frac{\gamma M m}{v^2} (1-3v^2) \left( \log \frac{M + \gamma m}{m + \gamma M}\right). \label{Nzf} \\
\tilde{N}_f^{x,\rm{tot}} & = 2(1+v^2)\frac{\gamma \tilde{E}Mm}{bv^4} \Bigg[ -(1-3v^2) \log \frac{M + \gamma m}{m + \gamma M} \nonumber \\ & \qquad +\left(v -\frac{8}{3}v^3 - (1-3v^2)\textrm{arctanh}\ \! v\right) \frac{M^2-m^2}{\tilde{E}^2} \Bigg]. \label{Nxf}
\end{align} 
Consistent to 2PM, this may be written as
\begin{align}
\tilde{N}^{z, \rm{tot}}_f & = -\tilde{N}^{z,\rm{tot}}_0 . \label{Nzf2} \\
\tilde{N}_f^{x, \rm{tot}} & =\chi \tilde{N}^{z,\rm{tot}}_0  - \frac{\Delta J }{2\gamma v} \frac{M^2-m^2}{M m}. \label{Nxf2}
\end{align}
We thus find that there is a change in the system's mechanical mass moment as a result of the scattering, an effect we will refer to as a ``scoot''.  Notice that there is a scoot at \textit{first} post-Minkowksian order (the logarithmic term in $\tilde{N}^{z,\rm{tot}}_f$), together with 2PM corrections (the remaining terms in $\tilde{N}^{z,\rm{tot}}_f$ and $\tilde{N}^{x,\rm{tot}}_f$).  We will discuss more details of this effect in a future publication \cite{gravscoot}. 

\section*{Acknowledgements}

It is a pleasure to acknowledge Leor Barack and Chia-Hsien Shen for helpful discussions.  This work was supported in part by NSF grant PHY1752809 to the University of Arizona.  

\appendix



\section{Mass evolution in the scalar theory}\label{app:mass}
The equations of motion for a scalar charge do not preserve the rest mass $m$, but rather involve an evolution equation \cite{Quinn:2000wa} analogous to the self-force \eqref{fq},
\begin{align}
    \frac{dm}{d\tau} = -q^2\int_{\rm -\infty}^{\tau^-} u^\mu \nabla_\mu G d\tau'.
\end{align}
With the approximations of Sec.~\eqref{sec:self-forces}, this becomes
\begin{align}
    \frac{dm}{d\tau} = - 2 q^2 (\mathcal{A}_1 + \xi \mathcal{B}_1),
\end{align}
where $\mathcal{A}_1$ and $\mathcal{B}_1$ are defined and given in Eqs.~\eqref{calA1} and \eqref{calB1}.  Denoting the value of the mass in the asymptotic past by $m_0$, the time evolution of the mass is
\begin{align}
    m(\tau) = m_0 - \frac{2q^2}{\gamma} \int_{-\infty}^{\gamma \tau} (\mathcal{A}_1 + \xi \mathcal{B}_1) dt'.
\end{align}
where the integral is over the background worldline parameterized by $t'$, i.e., $x=b$, $y=0$, $z=v t'$.  Using Eqs.~\eqref{calA1} and \eqref{calB1}, this integral is easily evaluated in closed form to give the full mass evolution $m(\tau)$.  A notable property of this function is that it returns to $m_0$ in the future, i.e.,
\begin{align}
    m(\tau \to \pm \infty) = m_0.
\end{align}
That is, there is no change in rest mass over the scattering process.

\section{Parameterized Trajectories}\label{sec:trajectories}

In this appendix we provide explicit formulae for the trajectory of the particle in the frame of the star, using the $t$ coordinate as a parameter.  For the gravitational case, the CEM-frame trajectories can be determined from these in the manner described in below Eq.~\eqref{t0} of the main text.  We distinguish the contributions to the motion from the various forces as follows:
\begin{align}
\mathbf{z}_{\rm scalar} & = (b + x_{g1} + x_q, 0, vt + z_{g1} +z_q) \label{zscalarfinal} \\
\mathbf{z}_{\rm EM} & = (b + x_{g1} + x_e, 0, vt + z_{g1} +z_e) \label{zEMfinal} \\
\mathbf{z}_{\rm grav} & = (b + x_{g1} + x_{g2} + x_m + x_{\rm mm}, 0, vt + z_{g1} + z_{g2} +z_m+z_{\rm mm}) \label{zgravfinal}
\end{align}
The subscript $q$ refers to the scalar self-force, the  subscript $e$ refers to the electromagnetic self-force, and the subscript $m$ refers to the gravitational self-force.  The subscript $\rm{mm}$ refers to the matter-mediated force, which acts at the same order as the gravitational self-force.  The subscriptions $g1$ and $g2$ refer to the geodesic terms (gravitational forces) at order $M$ and $M^2$, respectively.  
In presenting the results, we will evaluate all functions explicitly in terms of $t$, with the exceptions of
\begin{align}
r & = \sqrt{b^2+v^2t^2} \\
s & = b^2 + t^2 v^2 \gamma^{-2}.
\end{align}
Here $r$ is just the radius of the particle as measured in flat spacetime with the background straight-line motion, while $s$ is a positive quantity with no clear interpretation.   

The first-order geodesic terms are given by
\begin{align}
x_{g1} & = - \frac{M}{bv^2}(1+v^2) (r+vt) \label{xg1} \\
z_{g1} & = \frac{M(1-3v^2)}{ v^2}\textrm{arctanh}\frac{vt}{r}.  \label{zg1}
\end{align}
These were already provided in Eqs.~\eqref{zfull} and \eqref{xfull} (and again in $\eqref{zfullt}$ and \eqref{xfullt}), where the notation $z=vt$ was used.  The second-order geodesic terms follow from a straightforward calculation using the Schwarzschild metric in isotropic coordinates.  The results are\footnote{In Eq. \eqref{zg2}, we have corrected an error from the previous arXiv version. We thank Hongji Wei for helping find this error.}

\begin{align}
    x_{g2}(t) & = \frac{M^2}{8 b v^4} \Bigg(\frac{8 \left(v^2+1\right)^2 v t}{r} + 8  \left(v^2+1\right) \left(3 v^2-1\right) \log
   \frac{r+vt}{b}-\frac{3 \pi  v^3t}{b}(4+v^2)\nonumber \\ &+2
   \left(v^2-2\right)^2+\frac{8 \left(3 v^4+2 v^2-1\right)vt
   }{r}\textrm{arctanh}\frac{vt}{r}
   -\frac{6 \left(v^2+4\right) v^3t}{b} \textrm{arctan}\frac{vt}{b}\Bigg) \label{xg2} \\
    z_{g2}(t) & = \frac{M^2}{8bv^4}\Bigg(-\frac{8\left(v^2+1\right) \left(v^2t^2\left(v^2+1\right)+2 b^2 \gamma^{-2}\right)}{b r}-\frac{8\left(v^2+1\right)^2 vt}{b}\nonumber\\ &+\frac{8b \left(1-3
   v^2\right)^2 }{r}\textrm{arctanh}\frac{vt}{r}+ 6v^2\left(4-9v^2\right)\left(\textrm{arctan}\frac{vt}{b}+\frac{\pi}{2}\right)\Bigg).\label{zg2}
\end{align}

The self-force terms result from the integrals described in Sec.~\ref{sec:motion}.  The scalar results are

\begin{align}
    x_q(t) &= \frac{\gamma M q^2}{24 b^2 m} \Bigg(\frac{2 b^3 \left(-12 \xi +4 (3 \xi -1) v^2+3\right)}{s}-\frac{2 b  rt}{\gamma^2 s^3} \Bigg(\frac{2 b^2 v^2t^2}{\gamma^2}
    \left(-12 \xi +3 (4 \xi +1) v^2+v^4+4\right)\nonumber\\ &+b^4 \left(-12 \xi +(12 \xi +5) v^2-6 v^4+4\right)+\frac{v^4 t^4 }{\gamma^4} \left(-12 \xi +(12 \xi +1)
   v^2+4\right)\Bigg)\nonumber\\ &-\frac{6 t  \left((4 \xi -1) v^2-4 \xi \right)}{\gamma v}\left( \arctan\frac{
   r}{\gamma v b}+\arctan\frac{vt }{\gamma b}\right) -\frac{16 b^7 v^4}{s^3}+\frac{4 b^5 v^2 \left(3 v^2+4\right)}{s^2}\nonumber\\ &+2 b \left(-12 \xi +\frac{12 \xi
   }{v^2}-4 v-\frac{4}{v}+3\right)\Bigg) \label{xq} \\
    z_q(t) &= \frac{\gamma M q^2 }{24 b^2  m } \Bigg(\frac{2 b^2    \left(-12 \xi +4 (3 \xi -2) v^2+3\right)vt}{\gamma^2 s}- \frac{2 b^2r}{\gamma^2 vs^3 } \Big(\frac{2 b^2v^2t^2}{\gamma^4} \left(12 \xi +3 v^2-4\right)\nonumber\\ &+b^4 \left(12 \xi +(5-12 \xi ) v^2+2 v^4-4\right)-\frac{v^4t^4}{\gamma^4}
   \left(-12 \xi +3 (4 \xi -3) v^2+4\right)\Big)-\frac{16 b^6 v^5t}{\gamma^2 s^3}\nonumber\\ &+\frac{6b\left((4 \xi +1) v^2-4 \xi \right) }{\gamma v^2}\left( \arctan\frac{r}{\gamma v b}+\arctan\frac{vt}{\gamma b}\right)+\frac{4 b^4\left(v^2+4\right) v^3 t}{\gamma^2 s^2}\Bigg) \label{zq}
\end{align}

\begin{align}
    x_e(t) &= \frac{e^2 \gamma M}{24 b^2 m v^2} \Bigg( \frac{-4 b^5 v^4 \left(3 v^2+4\right)}{s^2}+\frac{2 b^3 v^2 \left(10 v^2+3\right)}{s}+\frac{16 b^7 v^6}{s^3}\nonumber\\ &-\frac{2 b  v^2t}{\gamma^2s^3} r \left(2 b^2 v^2 t^2 \left(v^6-4 v^4+v^2+2\right)+b^4 \left(6 v^4+v^2+2\right)+\frac{v^4 t^4 }{\gamma^4}
   \left(5 v^2+2\right)\right)\nonumber\\ &+\frac{6v t}{\gamma} \left(v^2+2\right) \left(\arctan\frac{r}{\gamma v b} + \arctan\frac{v t}{\gamma b}\right)-2 b \left(8 v^3-3 v^2+8 v-6\right)\Bigg) \label{xe} \\
    z_e(t) &= \frac{e^2 \gamma M }{24 b^2 m v^2} \Bigg(\frac{16 b^6  v^7 t}{\gamma^2 s^3}-\frac{4 b^4  \left(v^2+4\right)
   v^5 t}{\gamma^2 s^2}-\frac{2 b^2 \left(10 v^2-3\right) v^3 t}{ \gamma^2 s}\nonumber\\ &+\frac{2 v b^2r \left(2 b^2 v^2 t^2 \left(3 v^6+8 v^4-13 v^2+2\right)+b^4 \left(2 v^4-13 v^2+2\right)-\frac{v^4 t^4}{\gamma^4}
   \left(9 v^2-2\right)\right)}{\gamma^2 s^3}\nonumber\\ &+\frac{6  b \left(5 v^2-2\right) }{ \gamma }\left(\arctan\frac{
   r}{\gamma v b}+\arctan\frac{vt}{\gamma b}\right)\Bigg)\label{ze}
\end{align}

\begin{align}
    x_m(t) &= \frac{\gamma m M}{12 b^2 v} \Bigg(\frac{2 b^5 v^3 \left(3 v^2+4\right)}{s^2}-\frac{b^3 v \left(28 v^2+45\right)}{s}+b \left(44
   v^2-21 v+44\right) -\frac{8 b^7 v^5}{s^3} \nonumber\\&+ \frac{vt}{\gamma^2} \frac{r}{s^3} \Big(2 b^3 v^2 t^2\left(v^6-22 v^4-23 v^2+44\right)+b^5 \left(6
   v^4+19 v^2+44\right)+\frac{b v^4 t^4}{\gamma^4} \left(23 v^2+44\right)\Big)\nonumber\\ &-\frac{21 v^2t}{\gamma} \left(\arctan\frac{r}{\gamma v b}+\arctan\frac{t v }{\gamma b}\right)\Bigg) \label{xm} \\ 
    z_m(t) &= \frac{\gamma  m M}{12 b^2v} \Bigg(\frac{-8 b^6 v^6t}{\gamma^2 s^3}+\frac{2 b^4 \left(v^2+4\right) v^4t}{\gamma^2 s^2}-\frac{b^2  \left(45-64 v^2\right)v^2t}{\gamma^2 s}\nonumber\\& -\frac{b^2 r \left(2 b^2 v^2 t^2 \left(3 v^6+62 v^4-109
   v^2+44\right)+b^4 \left(2 v^4-67 v^2+44\right)-\frac{v^4t^4 }{\gamma^4} \left(63 v^2-44\right)\right)}{2\gamma^2  s^3}\nonumber\\& +\frac{69  v b}{\gamma}\arctan\frac{\gamma v \left(-t\gamma^{-2}  r+b^2\right)}{b \left(r+t v^2\right)}-\frac{69 \pi v
    b}{2 \gamma}\Bigg) \label{zm}
\end{align}

\begin{align}
    x_{\rm mm}(t) &= \frac{\gamma m M}{2 b^2 v^4} \Bigg(\frac{2 b^3 \left(3-v^2\right) v^4}{s}-2 b \left(3 v^4+2 v^2-1\right) \left(\log
   \frac{b}{r+vt} + \textrm{arctanh}(v)\right)\nonumber\\ &-\frac{2 b 
   (v-1) \left(b^2 \left(v^6+v^5-2 v^4-3 v^3+2 v^2+1\right)+t^2 v^2 \left(v^5-4 v^4-3
   v^3+v^2+1\right)\right)vt}{ r s}\nonumber\\ & -\frac{2 b  \left(3 v^4+2
   v^2-1\right) vt }{r}\textrm{arctanh}\frac{v \left(r-t\right)}{r-t
   v^2}-2 b \left(v^5-2 v^4+2
   v^3+v^2+v-1\right) \nonumber\\ & + \frac{ 2\left(v^2-3\right) v^3t}{\gamma} \left(\arctan\frac{vt}{\gamma b}+\arctan\frac{r}{\gamma v b}
   \right)\Bigg) \label{xmm}\\
    z_{\rm mm}(t) &= \frac{\gamma m M}{2b^2 v^4} \Bigg(\frac{2b^2  \left(v^4-4 v^2+3\right) v^5t}{s} +\frac{2 b^2 v^2t^2
   \left(-v^8+4 v^7-2 v^6-5 v^5+4 v^4+2 v^2+v-3\right)}{rs} \nonumber\\& +\frac{2b^4 \left(v^7+v^6-7 v^5+4 v^4+v^3+v^2+v-2\right)- 2v^4t^4
   \left(v^4-1\right)^2}{r s} \nonumber\\& +\frac{
   2\left(v^2-3\right) v^2 b}{\gamma}\arctan\frac{\gamma v \left(-t \gamma^{-2} r+b^2\right)}{b 
   \left(r+t v^2\right)} +\frac{2\left(1-3 v^2\right)^2 b^2 }{r}\textrm{arctanh}\frac{v
   \left(t-r\right)}{r-t v^2}\nonumber\\& + 2\left(v^2-1\right)
   \left(v^2+1\right)^2 vt-\frac{\pi  v^2 \left(v^2-3\right)b}{ \gamma }\Bigg)\label{zmm}
\end{align}

The $G^x$ and $G^z$ integrals defined in Sec.~\ref{sec:motion} are given in terms of these definitions by 
\begin{align}
    G^z_{\rm scalar}(t) & = \gamma^2 z_q(t) \label{Gzscalar} \\
    G^x_{\rm scalar}(t) & = x_q(t) \\
    G^z_{\rm EM}(t) & = \gamma^2 z_e(t)\\
    G^x_{\rm EM}(t) & =  x_e(t) \\
    G^z_{\rm grav}(t) & = \gamma^2( z_m + z_{\rm mm} )\\
    G^x_{\rm grav}(t) & = x_m + x_{\rm mm}. \label{Gxgrav}
\end{align}
 
\section*{References}

\bibliographystyle{utphys}
\bibliography{2pmscattering.bib}

\end{document}